\documentclass[
   aps,
   pra,
   twocolumn,
   english,
   superscriptaddress,
   floatfix,
]{revtex4-2}

\usepackage{
    times,
    bbold,
    bm,
    dsfont,
    amsmath,
    amsthm,
    amssymb
}

\usepackage{float}

\usepackage{physics}

\usepackage{booktabs}

\usepackage{color}
\usepackage[usenames, dvipsnames]{xcolor}
\usepackage{xkcdcolors}

\usepackage{xspace}

\usepackage{hyperref}
\hypersetup{
    pdftitle={Experimental realization of a photonic weighted graph state for quantum metrology},
    pdfauthor={Unathi Skosana, Byron Alexander, Changhyoup Lee, Mark Tame},
    citecolor=PineGreen,
    linkcolor=RedViolet,
    urlcolor=RedViolet,
    colorlinks=true,
    linktocpage=true,
    breaklinks=true,
    bookmarksopen=true,
    backref=true,
    linktoc=all,
}

\usepackage[newcommands]{ragged2e}
\usepackage{graphicx}
\graphicspath{{gfx/}}
\usepackage[export]{adjustbox}
\usepackage[justification=Justified,singlelinecheck=false,format=plain]{caption}
\usepackage[justification=Justified,singlelinecheck=false,format=plain]{subcaption}

\begin{document}

\title{Experimental realization of a photonic weighted graph state for quantum metrology}
\author{Unathi Skosana}
\email{ukskosana@gmail.com}
\affiliation{Department of Physics, Stellenbosch University, Matieland 7602, South Africa}
\author{Byron Alexander}
\affiliation{Department of Physics, Stellenbosch University, Matieland 7602, South Africa}
\author{Changhyoup Lee}
\affiliation{Department of Physics, Hanyang University, Seoul 04763, South Korea}
\author{Mark Tame}
\affiliation{Department of Physics, Stellenbosch University, Matieland 7602, South Africa}

\begin{abstract}
    \noindent
    Quantum metrology seeks to push the boundaries of measurement precision by
    harnessing quantum phenomena. Conventional methods often rely on maximally
    entangled resources, with states that are usually challenging to produce and
    sustain in practical setups. Here, we show that the maximally entangled
    constraint can be lifted by experimentally realizing a photonic two-qubit
    weighted graph state with an arbitrarily tunable graph weight. We use the
    generated state as a resource for quantum-enhanced phase sensing. We
    experimentally characterize the state and study its minimum estimator
    variance for two distinct local measurement bases as the graph weight
    varies from the maximally entangled to weakly entangled limit. We find
    excellent quantitative agreement with theoretical predictions, and observe a
    gain in precision beyond the classically attainable precision limit for
    graph weights substantially below the maximally entangled limit. This
    confirms that considerably less entanglement is required to achieve a
    quantum advantage. Albeit non-scalable in our test setup, this work
    represents the first experimental realization of weighted graph states with
    a tunable graph weight using linear optics. We expect more scalable
    versions of the model to be possible in an on-chip photonic platform.
\end{abstract}

\maketitle

\section{Introduction\label{sec:introduction}}
Graph states~\cite{Briegel2001,Raussendorf2001} are highly entangled quantum
states that serve as essential primitives for a wide range of quantum
information applications. Initially conceived as a canonical resource for
measurement-based universal quantum
computation~\cite{Raussendorf2003,Walther2005,Menicucci2006}, graph states have
since become crucial for many other applications. These include being used as
moderately sized resource states in fusion-based universal quantum
computation~\cite{Bartolucci2023,Thomas2024}, quantum communication protocols
such as entanglement purification~\cite{Dur2003}, secret
sharing~\cite{Markham2008}, and quantum‑enhanced
metrology~\cite{Friis2017,Shettell2020}. A major drawback is that the
realization of graph states is experimentally challenging. Among the physical
architectures on which graph states have been realized, which
include superconducting circuits~\cite{Song2017,Mooney2019}, individually
addressable single-atoms~\cite{Thomas2022,Thomas2024}, trapped
ions~\cite{Lanyon2013,Pogorelov2021,Moses2023,Ringbauer2025,Kang2025}, neutral
atoms~\cite{Mandel2003}, nitrogen-vacancy centers~\cite{Cramer2016}, and
quantum dots~\cite{Istrati2020,Cogan2023,Huet2025}, photonic architectures
offer many advantages for quantum information: they are operational at room
temperature, possess long coherence times, and provide a natural substrate for
quantum communication~\cite{Obrien2007,Wang2019,Zhong2020}.

The generation of photon-photon entanglement for creating quantum resources
such as graph states and carrying out quantum information processing utilizes
several key approaches. These include probabilistic photon-photon entanglement
generation schemes that use only linear optical
elements~\cite{Browne2005,Gao2010,Li2020}, and those that additionally rely on
single-photon sources based on spontaneous parametric down‑conversion
(SPDC)~\cite{Walther2005,Park2007,Lee2012,Wang2018,Zhong2018} to achieve higher
success rates than the former schemes, and very-large-scale-integration
technologies that combine both schemes to achieve miniaturization, modularity
and scalability~\cite{Bao2023,Pont2024,Wang2025,Yang2025}. Alternatively,
light-matter interaction techniques, such as those leveraging highly coupled
cavity-QED systems and emitter-based schemes, can enable deterministic
entanglement
generation~\cite{Firstenberg2013,Firstenberg2016,Tiarks2016,Thompson2017,Tiarks2018,Sagona2020,Ferreira2024}.

The generation procedure of graph states in particular requires the faithful
realization of pairwise maximally entangling controlled-Z (CZ) gates to yield
high-fidelity states. The controlled-Z gate is rarely experimentally realized
with unit gate fidelity, and often results in a pairwise entangling gate that
implements a controlled‑phase gate with arbitrary phase in
photonic~\cite{Young2011,Firstenberg2013,Firstenberg2016,Tiarks2016,Thompson2017,Tiarks2018,Sagona2020}
and other physical systems~\cite{Qin2025,OSullivan2025}. When the controlled-Z
gates in the generation procedure of graph states are replaced by
controlled-phase gates, with the value of phases representing weights of the
edges between vertices, the resulting quantum states are called weighted graph
states~\cite{Dur2005,Hein2006,Hartmann2007}. Such quantum states can be seen as
a generalization of standard graph states, where the $\pi$ phase due to the
controlled-Z gates between the graph vertices is replaced by an arbitrary phase
value $\phi$ due to controlled-phase gates.

Weighted graph states inherit many of the properties of standard graph states,
while their generality broadens their utility across quantum information
science. A few select examples include the use of weighted graph states from a
computational standpoint as a variational ansatz for approximating the ground
state of strongly interacting spin systems such as Ising
models~\cite{Anders2006,Anders2007}. Furthermore, weighted graph states can be
leveraged in measurement-based quantum computation for generating random
quantum circuits~\cite{Plato2008} and the realization of a measurement-based
Toffoli gate~\cite{Tame2009}, efficient polynomial-time fidelity estimation
protocols~\cite{Hayashi2019}, and the detection of transitions in variable-range
interaction spin models~\cite{Ghost2024}. In fusion-based quantum computation,
the use of weighted graph states can lead to a higher success probability for
the fusion process~\cite{Rimock2024}. Weighted graph states have also been
proposed as a direct resource for use in quantum metrological applications such
as quantum sensing~\cite{Xue2012,Alexander2025}. This is in contrast to
treating weighted graph states as an intermediary and imperfect resource state
to produce standard graph states, as done in entanglement concentration
protocols~\cite{Paunkovi2002,Zhao2003,Hwang2007,Xiong2011,Sheng2012,Deng2012,Zhou2012,Choudhury2013,Li2014,Frantzeskakis2023}.

\newpage
\noindent
The inherent tunability of weighted graph states via the tunable graph weights
can be leveraged to achieve high-precision measurement of a physical parameter
beyond classical limits. When utilizing sub-classes of weighted graph states
for quantum sensing it is possible to maintain a quantum advantage for
non-maximal graph weights~\cite{Alexander2025}. This enables the systematic
balancing of a trade-off between entanglement and precision under practical
constraints, either due to imperfect or restricted experimental apparatus. The
trade-off allows one to identify regimes where reduced entanglement is
sufficient for an enhanced metrological performance, this represents a distinct
advantage over standard graph states with a fixed topology and graph weights.
Despite the compelling evidence from the theoretical framework establishing the
robustness of weighted graph states for metrological applications in
Ref.~\cite{Alexander2025} and the complementary strengths of photonic
architectures, to date no experimental work exists utilizing photonic weighted
graph states as a direct resource for quantum metrology.

In this study, we demonstrate the experimental realization of a photonic
two-qubit weighted graph state with a tunable graph weight using linear optics
and a single polarization-entangled photon-pair source based on SPDC. We use
the photonic weighted graph state for the metrological task of sensing a
parameter $\theta$ associated with a unitary transformation $U(\theta)$ to the
highest precision possible. Here, we vary the graph weight between $\pi$
(standard two-qubit graph state) and $0$ (product state), and show sensing
precision beyond the classical limit for weights substantially smaller than
$\pi$; experimentally confirming that considerably less entanglement is
sufficient for achieving a quantum advantage. This represents the first direct
demonstration of a weighted graph state with a tunable weight used in a
photonic quantum sensing application, bridging theory and experiment to
establish photonic weighted graph states as a practical resource for
quantum-enhanced metrology. Our work constitutes a proof-of-concept
demonstration, and while utilizing a non-scalable optical architecture, the
generation procedure presented here has the potential to be compatible with
scalable
platforms~\cite{Young2011,Firstenberg2013,Firstenberg2016,Tiarks2016,Thompson2017,Tiarks2018,Sagona2020}.

The paper is organized as follows. In Sec.~\ref{sec:experimental_setup} we
describe our experimental setup and explain implementation details.
Sec.~\ref{sec:state_characterization} describes the experimental results
concerning state characterization. Then, in Sec.~\ref{sec:quantum_sensing} we
describe the phase sensing protocol and experimental results. Finally, in
Sec.~\ref{sec:conclusion}, we give a summary and discuss future work.

\section{Experimental setup\label{sec:experimental_setup}}
\begin{figure*}[t!]
    \centering
    \vspace{-3ex}
    \includegraphics[width=.9\linewidth]{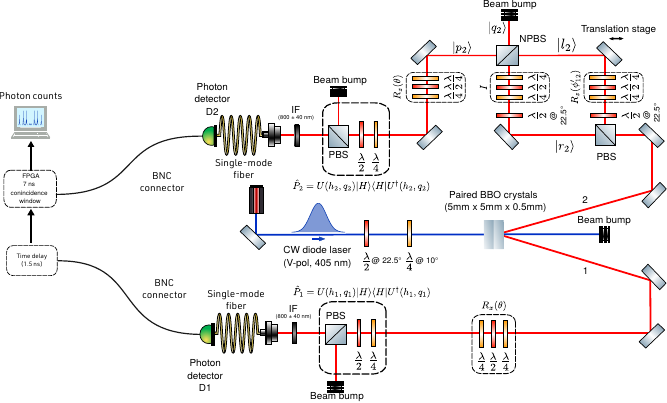}
    \caption{
	Experimental setup for creating the two-qubit weighted graph
	$\ket*{\Gamma_{\phi_{12}}}$ with a tunable graph weight, and for
	performing a quantum metrology protocol. A 405 nm continuous wave (CW)
	diode laser pump beam (vertically polarized) passes through a half-wave
	plate (HWP) set to $22.5^{\circ}$, a quarter-wave plate (QWP) set to
	$10^{\circ}$, and two cascaded $\beta$-barium borate crystals (5mm by
	5mm by 0.5mm each) with mutually perpendicular optic axes, to create
	the polarization-entangled $\ket*{\Phi^{(-)}}$ state. The generated
	photons are separated: photon 1 goes to a sensing stage. Photon 2 is
	split by a PBS into paths $r_2$ and $l_2$, forming a Mach-Zehnder
	interferometer (MZI). The $R_z(\phi'_{12})$ rotation in path $l_2$
	(implemented by a QWP-HWP-QWP configuration) is the core tuning
	mechanism for the graph weight $\phi_{12}$. The paths recombine at a
	non-polarizing beam splitter (NPBS), where port $p_2$ is selected. Both
	photons then enter the quantum sensing stage, where the phase $\theta$
	is encoded by the unitary $U(\theta) = R^{(1)}_x(\theta) \otimes
	R^{(2)}_z(\theta)$. Finally, both photons are detected after passing
	through an arbitrary polarization projective measurement (QWP-HWP-PBS),
	interference filters (800 nm $\pm$ 40 nm), and coupled into single-mode
	fibers for detection by avalanche photodiode (APD) single-photon
	detectors and a coincidence counting module. The detector output for
	photon 1 is delayed by 1.5 ns before it enters the coincidence counting
	module.
    }
    \label{fig:experiment_schematic}
\end{figure*}

Our experimental scheme in Fig.~\ref{fig:experiment_schematic} for generating a
weighted graph state with a tunable graph weight is based on the scheme reported
in Ref.~\cite{Park2007}. The scheme consists of a polarization-entangled
photon-pair source that generates the state $\ket*{\Phi^{(-)}} = (\ket*{H_1, H_2} -
\ket*{V_1, V_2})/\sqrt{2}$. Here, $H$ ($V$) denotes horizontal (vertical)
polarization, and the subscripts $1, 2$ refer to the photons, respectively. Our
photon-pair source consists of a 405 nm continuous wave diode laser (vertically
polarized) as the pump beam, a half-wave plate (HWP) set to $22.5^{\circ}$, a
quarter-wave plate (QWP) set to $10^{\circ}$, and two cascaded $\beta$-barium
borate (BBO) crystals with mutually perpendicular optic axes~\cite{Kwiat1999}.
A non-collinear type-I SPDC process in the cascaded BBO crystals generates
frequency-degenerate photon pairs at a wavelength of 810 nm, emitted into a
light cone with a half opening angle of $3^{\circ}$ relative to the pump beam
direction. The photon-pair source sends photon $1$ to the sensing and
measurement step described later. Photon $2$ goes through a series of optical
elements to generate the weighted graph state. First, it goes into a polarizing
beam splitter (PBS), which transmits (reflects) a horizontally (vertically)
polarized photon, and transforms the two-photon state $\ket*{\Phi^{(-)}} =
(\ket*{H_1, H_2} - \ket*{V_1, V_2})/\sqrt{2}$ to
\begin{align}
    \ket*{\Psi_\text{I}} = (\ket*{H_1, r_2, H_2} - \ket*{V_1, l_2, V_2})/\sqrt{2},
    \label{eq:standard_ghz}
\end{align}
\noindent
where $r_2$ ($l_2$) refers to the transmitted (reflected) paths for photon $2$,
respectively, as shown in Fig.~\ref{fig:experiment_schematic}. Photons in both
$l_2$ and $r_2$ go through a HWP set to $22.5^{\circ}$ with respect
to the fast axis and the vertical axis, which maps $\ket*{H} \mapsto \ket*{D}$ in
$r_2$ and $\ket*{V} \mapsto \ket*{A}$ in $l_2$, where $\ket*{D/A} =
(\ket*{H} \pm \ket*{V})/\sqrt{2}$. This is followed by a configuration of
QWP-HWP-QWP that implements a $R_{z}(\phi'_{12})$
rotation for photons in $l_2$ and an identity $I$ operator in $r_2$,
respectively. The $R_z(\phi'_{12})$ rotation on photons in $l_2$ is the core
tuning mechanism for the graph weight. A QWP-HWP-QWP configuration can
uniquely decompose a general $\mathrm{SU}(2)$
operator~\cite{Simon1990,Reddy2014,Passos2020}. After these operations, the state in
Eq.~\ref{eq:standard_ghz} becomes
\begin{align}
    &\ket*{\Psi_\text{II}} = (\ket*{H_1, r_2, D_2} - e^{-i\phi'_{12}/2}\ket*{V_1, l_2, D_{2}^{\pi + \phi'_{12}}})/\sqrt{2},&&
    \label{eq:standard_ghz_1}
\end{align}
\noindent
where $\ket*{D^{\pi + \phi'_{12}}} = (\ket*{H}+ e^{i(\pi +
\phi'_{12})}\ket*{V})/\sqrt{2}$. When photons from $l_2$ and $r_2$ combine on the input
ports of the non-polarizing beam splitter (NPBS), we obtain
\begin{align}
    &\ket*{\Psi_\text{III}} =(e^{-i\varphi_{r_2}} (\ket*{H_1, p_2, D_2} + \ket*{H_1, q_2, D_2}) - \nonumber \\
    & e^{-i\varphi_{l_2}}e^{-i\phi'_{12}/2} (\ket*{V_1, p_2, D_{2}^{\pi + \phi'_{12}}} - \ket*{V_1, q_2, D_{2}^{\pi + \phi'_{12}}}))/2. &&
    \label{eq:standard_ghz_2}
\end{align}
\noindent
Here, $p_2$ and $q_2$ are the output ports of the non-polarizing beam splitter,
defined by the transformations $\ket*{r_2} \to (\ket*{p_2} + \ket*{q_2})/\sqrt{2}$
and $\ket*{l_2} \to (\ket*{p_2} - \ket*{q_2})/\sqrt{2}$. Moreover,
$e^{-i\varphi_{r_2}}$  and $e^{-i\varphi_{l_2}}$ are relative phases acquired
by the optical paths $r_2$ and $l_2$, respectively. We take the detected
photons to be from the output port $p_2$, giving the state
\begin{align}
    &\ket*{\Psi_\text{IV}} = (\ket*{H_1, D_2} - e^{i\varphi^{\prime}}e^{-i\phi'_{12}/2}\ket*{V_1, D_{2}^{\pi + \phi'_{12}}})/\sqrt{2},
    \label{eq:standard_ghz_3}
\end{align}
\noindent
where we have collected all the $p_2$ terms, set $\varphi^{\prime} =
\varphi_{r_2} - \varphi_{l_2}$ and dropped a global phase. Denoting $\phi_{12}
= \pi + \phi'_{12}$ and $\varphi= \varphi^{\prime} -\phi'_{12}/2 + \pi$, the
expression in Eq.~\ref{eq:standard_ghz_3} becomes
\begin{align}
    \ket*{\Psi_\text{V}} = (\ket*{H_1, D_2} + e^{i\varphi}\ket*{V_1, D_{2}^{\phi_{12}}})/\sqrt{2}.
    \label{eq:weighted_graph_state_with_mzi_phase}
\end{align}
\noindent
When the phase difference $\varphi^{\prime}$ between the optical path $l_2$ and
$r_2$ becomes equal to $\phi_{12}'/2 - \pi$ from translating the mirror on the
stage, then $\varphi$ becomes equal to zero and we obtain a two-qubit weighted
graph state with the graph weight $\phi_{12}$:
\begin{align}
    \ket*{\Gamma_{\phi_{12}}} = (\ket*{H_1, D_2} + \ket*{V_1, D_{2}^{\phi_{12}}})/\sqrt{2}.
    \label{eq:weighted_graph_state}
\end{align}
\noindent
In the computational basis, where $\{\ket{0},\ket{1}\}$ is represented by $\{\ket{H}, \ket{V}\}$,
we have
\begin{align}
    \ket{\Gamma_{\phi_{12}}} = \mathrm{CZ}^{\phi_{12}}\ket{+}_1\ket{+}_2,
    \label{eq:standard_weighted_graph_stte}
\end{align}
\noindent
where $\mathrm{CZ}^{\phi_{12}} = \mathrm{diag}(1, 1, 1, e^{i\phi_{12}})$,
giving the standard definition of a weighted graph state. To encode a phase
parameter for quantum metrology, photon $1$ ($2$) goes through a QWP-HWP-QWP
configuration that implements a $R_x(\theta)$ ($R_z(\theta)$) rotation,
respectively. The phase encoding unitary operator $U(\theta) =
R^{(1)}_x(\theta)\otimes R^{(2)}_z(\theta)$ is relevant in the phase sensing
protocol described later. The rotation $R_{x}^{(1)}(\theta)$ is equivalent to a
phase rotation up to Hadamard operations, i.e., $R_{x}^{(1)}(\theta) = H
R_{z}^{(1)}(\theta)H$. In the initial state generation step, both sets of
waveplates are configured to implement the identity operator.

\noindent
The waveplate angles for each rotation and the identity operator can be found
in Appendix~\ref{sec:single_q_rotation_gates_with_pol_optics}. The measurement
steps ($\hat{P}_1$ and $\hat{P}_2$ in Fig.~\ref{fig:experiment_schematic}) for
the photon pairs are identical. Each photon goes through a QWP, HWP, and PBS,
where we only collect the transmitted photons. This combination of optics is
capable of realizing an arbitrary polarization projective measurement. Each
photon then passes through 800 nm $\pm$ 40 nm interference filters. Finally, the photons are coupled into
single-mode fibers. The single-mode fibers are then directly coupled into
silicon avalanche photodiode (APD) single-photon detectors. The detector output
for photon $2$ goes directly into a coincidence counting
module~\cite{Branning2009} with a coincidence window of 7 ns. The detector
output for photon $1$ is electronically delayed by 1.5 ns before entering the
coincidence counting module to compensate for the longer optical path of photon
$2$. This ensures both photons arrive within a single coincidence window.

\section{State characterization\label{sec:state_characterization}}

\subsection{Phase stability and control\label{subsec:phase_stability_and_control}}
The experimental setup described in the preceding section generates the target
two-qubit weighted graph state, $\ket*{\Gamma_{\phi_{12}}}$, with a tunable
graph weight $\phi_{12}$. Here, we characterize various aspects of the state
generation. In the first instance, we characterize the control of phase
difference, $\varphi^{\prime} = \varphi_{r_2} - \varphi_{l_2}$, inside the
Mach-Zehnder interferometer (MZI). This phase difference is critical for state
generation; achieving the condition $\varphi^{\prime}=\varphi_{12}/2 - \pi$ is
necessary to realize the desired weighted graph state. We quantify the
experimental capability to distinguish between different target values of
$\varphi^{\prime}$ using fringe visibility measurements. To do this, we perform
fringe visibility measurements on an intermediate two-photon state,
$(\ket*{H_1, H_2} - e^{i\varphi^{\prime}}\ket*{V_1, H_2})/\sqrt{2}$. In the
experiment, this state is obtained from Eq.~\ref{eq:standard_ghz} by setting
the HWPs inside $l_2$ and $r_2$ to $45^{\circ}$ and $0^{\circ}$, respectively.
The $R_z(\phi'_{12})$ rotation in $l_2$ is configured to implement the identity
operator. When the polarization measurement optics project $\hat{P}_{1} \otimes
\hat{P}_{2} = \ket*{A_1}\bra*{A_1}$ $ \otimes \ket*{H_2}\bra*{H_2}$, we obtain
the following detection probability:
\begin{align}
    P(o_1 = A_{1}, o_2 = H_{2}) = (1 + \cos{\varphi^{\prime}})/2.
    \label{eq:visibility}
\end{align}
\noindent
From Eq.~\ref{eq:visibility}, we see that we can perform fringe visibility
measurements by translating the mirror in $l_2$ to sweep the value of
$\varphi^{\prime}$ over a full cycle $[0, 2\pi)$ and recording the coincidence
photon counts. We then take the maximum and minimum coincidence counts
recorded, and calculate the fringe visibility as $V = (N_\text{max} -
N_\text{min})/(N_\text{max} + N_\text{min})$, where $N_\text{max}
(N_\text{min})$ is the maximum (minimum) photon count.

\begin{figure}[H]
    \centering
    \includegraphics[width=\linewidth]{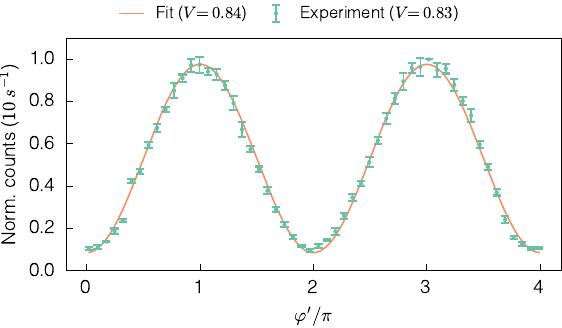}
    \caption{
	Normalized photon counts and fit as a function of the phase difference
	between the two optical paths $\varphi^{\prime}$, varied by translating
	the mirror on a stage in steps of 60 nm. Each data point is an average
	of 10 measurements, where each measurement is from accumulated photon
	counts over a 10-second interval. The normalized photon counts are
	fitted to $f(x) = a\cos\left(b x + c\right) + d$, where the parameters
	of the fit are $a, b, c$ and $d$. The error bars represent the standard
	deviation of the mean. The discontinuity around the 6th step was due to
	a double jog. See Appendix~\ref{sec:fringe_visibility} for more
	details.
    }
    \label{fig:bell_state_fv_ah_fit}
\end{figure}

\newpage
\noindent
Fig.~\ref{fig:bell_state_fv_ah_fit} shows normalized photon counts as we vary
$\varphi^{\prime}$, where each data point is an average of 10 measurements,
each collected over a 10-second interval, with a maximum of 150 coincidence
counts per second. We achieve a fringe visibility of $V=83\%$, confirming that
the phase $\varphi^{\prime}$ due to the optical path length difference is
stable enough to be reliably controlled. See
Appendix~\ref{sec:fringe_visibility} for more details.

\subsection{Intermediate state tomography\label{sec:intermediate_state_tomography}}
Next, we verify the fidelity~\cite{MikeIke2011} of the generated state via
quantum state tomography (QST). Here, the primary objective is to reconstruct
the density matrix for the intermediate polarization-entangled state
$(\ket*{H_1, H_2} - e^{i\varphi^{\prime}}\ket*{V_1, V_2})/\sqrt{2}$ at
specific targeted values for the phase $\varphi^{\prime}$. This state is
obtained by setting the HWPs inside $l_2$ and $r_2$ both to $0^{\circ}$. The
$R_z(\phi'_{12})$ rotation in $l_2$ is configured to implement the identity
operator. Reconstructing the two-qubit density matrix requires performing a set
of 16 different, non-redundant polarization projective
measurements~\cite{James2001}. The measurement bases are selected from the
product of the three standard single-photon bases: rectilinear basis
$\{\ket*{H}, \ket*{V}\}$, diagonal basis $\{\ket*{D},\ket*{A}\}$ and circular
basis $\{\ket*{R}, \ket*{L}\}$. These measurements are realized experimentally
by the QWP, HWP, and PBS configuration in the measurement stage, as described
in the preceding section.

For each of the 16 measurement settings we record coincidence counts over a
10-second interval. The waveplate angles corresponding to each polarization
measurement basis can be found in
Appendix~\ref{sec:polarization_measurement_bases}. The experimentally
determined counts are processed using a maximum likelihood estimation (MLE)
algorithm that reconstructs a valid physical density matrix. In our experiment,
we choose to characterize the state when $\varphi^{\prime}$ takes on the values $\varphi'= 0$,
$\pi/4$, $\pi/2$, $3\pi/4$, and $\pi$. To set $\varphi^{\prime}$ to one of
these values, we first monitor the photon counts for the projector $\hat{P}_1
\otimes \hat{P}_2 = \ket*{D_1}\bra*{D_1}$ $ \otimes \ket*{D_2}\bra*{D_2}$. For the
intermediate state $(\ket*{H_1, H_2} - e^{i\varphi^{\prime}}\ket*{V_1,
V_2})/\sqrt{2}$, this projector has the following detection probability,
\begin{align}
    P(o_1 = D_{1}, o_2 = D_{2}) = (1 - \cos{\varphi^{\prime}})/2.
    \label{eq:visibility_2}
\end{align}
\noindent
Thus, targeting the phases $\varphi^{\prime} = 0\text{ and }\pi$ corresponds to
cases where the detection probability above takes on a minimum and a maximum
value, respectively. The intermediate phases are then targeted by translating
the mirror inside path $l_2$ and recording the total number of discrete steps
it takes to go from the minimum to the maximum number of photon counts. Then,
starting from a minimum, $\pi/4$ corresponds to taking $25\%$ of the total
number of steps from minimum to maximum, $\pi/2$ corresponds to $50\%$ of the
total number of steps, and so forth. While performing projective measurements,
the phase $\varphi^{\prime}$ can drift from one value to another. We correct
for this drift by using a `dither lock' approach, where we keep track of photon
counts from fixed projective measurements and use them as checkpoints in
between tomography projective measurements for a particular phase. If there is
a drift, we actively correct for it by translating the mirror in $l_2$ until we
recover the checkpoint value.


\begin{figure}[H]
    \centering
    \begin{subfigure}[T]{\linewidth}
	\caption{\label{fig:s8_min_dd_real}}
	\includegraphics[width=\linewidth]{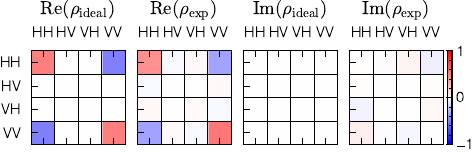}
    \end{subfigure}
    \par\bigskip
    \begin{subfigure}[T]{\linewidth}
	\caption{\label{fig:s8_min_14_dd_real}}
	\includegraphics[width=\linewidth]{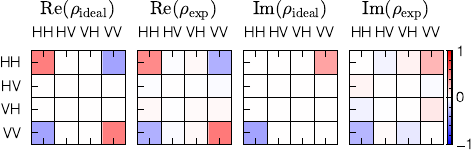}
    \end{subfigure}
    \par\bigskip
    \begin{subfigure}[T]{\linewidth}
	\caption{\label{fig:s8_min_24_dd_real}}
	\includegraphics[width=\linewidth]{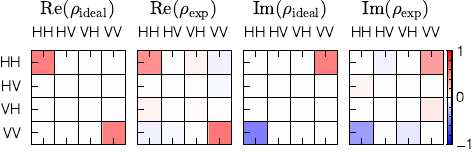}
    \end{subfigure}
    \par\bigskip
    \begin{subfigure}[T]{\linewidth}
	\caption{\label{fig:s8_min_34_dd_real}}
	\includegraphics[width=\linewidth]{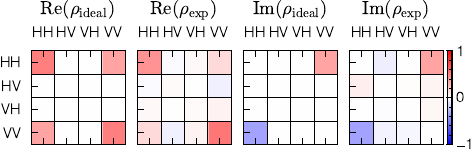}
    \end{subfigure}
    \par\bigskip
    \begin{subfigure}[T]{\linewidth}
	\caption{\label{fig:s8_max_dd_real}}
	\includegraphics[width=\linewidth]{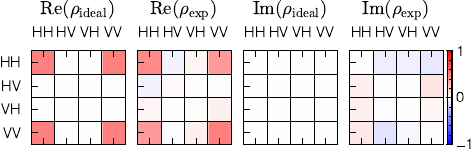}
    \end{subfigure}
    \caption{
	Real and imaginary parts of the reconstructed density matrices for the
	state $(\ket*{H_1, H_2} -e^{i\varphi^{\prime}}\ket*{V_1,
	V_2})/\sqrt{2}$ where $\varphi^{\prime}$ takes on the values
	\textbf{(a)} $\varphi^{\prime}=0$, \textbf{(b)}
	$\varphi^{\prime}=\pi/4$, \textbf{(c)} $\varphi^{\prime}=\pi/2$,
	\textbf{(d)} $\varphi^{\prime}=3\pi/4$, and \textbf{(e)}
	$\varphi^{\prime}=\pi$, respectively. All the density matrices are
	obtained using a maximum likelihood estimation (MLE) that takes the
	photon counts collected over a 10-second interval for each of the 16
	projective measurements in Tab.~\ref{tab:2q_tomo_measurements}.
    }
    \label{fig:bell_state_den_mat_s8}
\end{figure}

\newpage
\noindent
Fig.~\ref{fig:bell_state_den_mat_s8} shows reconstructed density matrices for
the intermediate state $(\ket*{H_1, H_2} -e^{i\varphi^{\prime}}\ket*{V_1,
V_2})/\sqrt{2}$ for the phases $\varphi^{\prime}=0$, $\pi/4$, $\pi/2$,
$3\pi/4$, and $\pi$. On average, the fidelities for each state were found to be
$F(\rho_\text{ideal},\rho_\text{exp}) = 0.8456 \pm 0.0140$, $0.9041 \pm
0.0153$, $0.8635 \pm 0.0130$, $0.8335 \pm 0.0190$, and $0.8754 \pm 0.0184$,
respectively.

\subsection{Weighted graph state tomography\label{subsec:weighted_graph_state_tomography}}
We now reconstruct the density matrices for the weighted graph state in
Eq.~\ref{eq:weighted_graph_state} for specific targeted values of the graph
weight $\phi_{12}$. This state is obtained by setting the HWPs in $l_2$ and
$r_2$ to $22.5{^\circ}$, and for a graph weight $\phi_{12}$, the
$R_{z}(\phi'_{12})$ rotation in $l_2$ is configured such that $\phi'_{12} =
\phi_{12} - \pi$. Moreover, for each graph weight $\phi_{12}$, we have to
ensure that the relative phase difference $\varphi'$ between the paths $l_2$
and $r_2$ in the MZI in
Fig.~\ref{fig:experiment_schematic} equals $\phi'_{12}/2 - \pi$, such that the
relative phase $\varphi$ in Eq.~\ref{eq:weighted_graph_state_with_mzi_phase}
equals zero. This is done by monitoring the photon counts for the projector
$\hat{P}_1 \otimes \hat{P}_2 = \ket*{D_1}\bra*{D_1}$ $ \otimes
\ket*{H_2}\bra*{H_2}$. The detection probability for this projective
measurement is given by,
\begin{align}
    P(o_1 = D_{1}, o_2 = H_{2}) &= (1 + \cos{\varphi})/2.
    \label{eq:visibility_3}
\end{align}
\noindent
The above detection probability reaches a maximum when the aforementioned
condition $\varphi' = \phi'_{12}/2 - \pi$ is satisfied. This is the required
condition for realizing a weighted graph state with weight $\phi_{12}$ in
Eq.~\ref{eq:weighted_graph_state}. Thus, for a fixed weight $\phi_{12}$, we
begin by fixing the rotation angle of the $R_{z}(\phi'_{12})$ rotation to
$\phi'_{12}=\phi_{12} + \pi$ and vary $\varphi'$ by translating the mirror in
$l_2$ until we observe a maximum number of coincidences for the projective
measurement, which effectively sets $\varphi=0$ and consequentially
$\varphi'=\phi'_{12}/2 - \pi$. For our reconstruction, we target the weighted
graph state when the graph weights $\phi_{12}$ take on the values $\phi_{12} = 0$, $\pi/4$,
$\pi/2$, $3\pi/4$ and $\pi$. Similarly here, when performing projective
measurements we checkpoint the maximum photon counts and correct for any drift
by translating the mirror inside $l_2$ until we recover the checkpoint value.

\begin{figure}[H]
    \centering 
    \begin{subfigure}[T]{\linewidth}
	\caption{\label{fig:s10_weight_zero_pi_real}}
	\includegraphics[width=\linewidth]{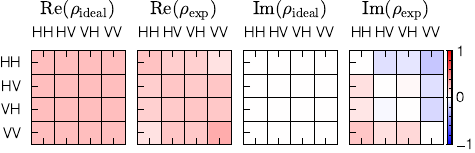}
    \end{subfigure}
    \par\bigskip
    \begin{subfigure}[T]{\linewidth}
	\caption{\label{fig:s10_weight_34_pi_real}}
	\includegraphics[width=\linewidth]{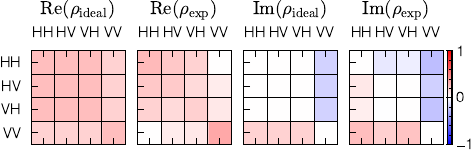}
    \end{subfigure}
    \par\bigskip
    \begin{subfigure}[T]{\linewidth}
	\caption{\label{fig:s10_weight_24_pi_real}}
	\includegraphics[width=\linewidth]{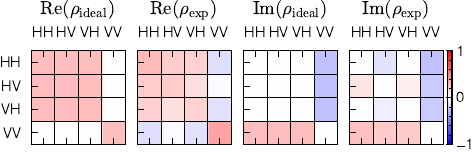}
    \end{subfigure}
    \par\bigskip
    \begin{subfigure}[T]{\linewidth}
	\caption{\label{fig:s10_weight_14_pi_real}}
	\includegraphics[width=\linewidth]{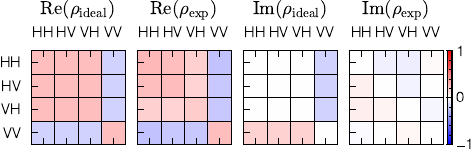}
    \end{subfigure}
    \par\bigskip
    \begin{subfigure}[T]{\linewidth}
	\caption{\label{fig:s10_weight_pi_real}}
	\includegraphics[width=\linewidth]{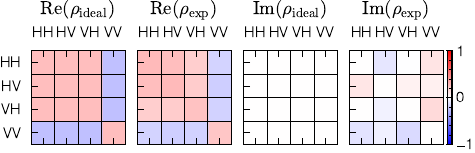}
    \end{subfigure}
    \caption{
	Real and imaginary parts of the reconstructed density matrices for the
	weighted graph states $(\ket*{H_1, D_2} + \ket*{V_1,
	D_{2}^{\phi_{12}}})/\sqrt{2}$ where $\phi_{12}$ takes on the values
	\textbf{(a)} $\phi_{12}=0$, \textbf{(b)} $\phi_{12}=\pi/4$,
	\textbf{(c)} $\phi_{12}=\pi/2$, \textbf{(d)} $\phi_{12}=3\pi/4$, and
	\textbf{(e)} $\phi_{12}=\pi$, respectively.  All the density matrices
	are obtained using a maximum likelihood estimation (MLE) that takes the
	photon counts collected over 10 seconds interval for each of the 16
	projective measurements in Tab.~\ref{tab:2q_tomo_measurements}.
    }
    \label{fig:graph_state_s10}
\end{figure}

Fig.~\ref{fig:graph_state_s10} shows reconstructed density matrices for the
weighted graph states $\ket*{\Gamma_{\phi_{12}}}$ where the weights take on the
values $\phi_{12}=0$, $\pi/4$, $\pi/2$, $3\pi/4$, and $\pi$. On average, the
fidelities for each state were found to be
$F(\rho_\text{ideal},\rho_\text{exp}) = 0.8133 \pm 0.0156$, $0.8134 \pm
0.0190$, $0.8302 \pm 0.0178$, $0.7943 \pm 0.0128$ and $0.8349 \pm 0.0104$,
respectively. When the graph weight $\phi_{12}$ takes on the value $\pi$, we
recover the standard maximally-entangled two-qubit graph state, i.e.,
$\ket*{\Gamma_{\pi}} = (\ket*{H_1, D_2} + \ket*{V_1, A_2})/\sqrt{2}$, our
reconstructed density matrix at this weight has a concurrence of $C=0.780 \pm
0.229$, indicative of the presence of
entanglement~\cite{Hill1997,Romero2007,Horodecki2009}. As we decrease the
weight towards zero, the concurrence of the reconstructed density matrices
decreases all the way down to $C=0.0895 \pm 0.0152$, indicative of a state close
to a separable state. Indeed, the graph weight $\phi_{12}=0$ leads to the ideal
product state $\ket*{\Gamma_{0}} = \ket*{D_1}\otimes\ket*{D_2}$, which has a
concurrence of zero.

\clearpage

\section{Quantum sensing\label{sec:quantum_sensing}}

\subsection{Phase encoding\label{sec:phase_encoding}}
Having characterized the generated weighted graph states
$\ket*{\Gamma_{\phi_{12}}}$ for a range of graph weights, we study utilizing
them as a resource for quantum metrology. Here, we leverage the generated
states to probe a phase $\theta$ with the primary objective of achieving the
maximum measurement precision offered by the state at each graph weight
$\phi_{12}$. Similar to Ref.~\cite{Alexander2025}, the scalar parameter
$\theta$ is encoded in the state $\rho=\ketbra*{\Gamma_{\phi_{12}}}$ by a
unitary transformation $U(\theta)$ due to a Hamiltonian $\hat{H}$
\begin{align}
    \rho \mapsto \rho_{\theta} := U(\theta)\rho U^{\dagger}(\theta),\> \text{ where } U(\theta) = e^{-i\hat{H}\theta/\hbar}.
\end{align}
\noindent
The unitary operator represents the sensing process, and the parameter
$\theta$ dictates the strength or duration of an interaction represented by a
Hamiltonian operator $\hat{H}$. Given this, we seek to determine the estimator
variance $(\Delta \theta)^2$ for an unbiased estimator $\hat{\theta}$ of
$\theta$,
\begin{align}
    (\Delta \theta)^2 = \frac{(\Delta A)^2}{\abs*{\partial_{\theta}\expval*{\hat{A}}}^2},
    \label{eq:estimator_variance_for_theta}
\end{align}
\noindent
where $(\Delta A)^2 := \expval*{\hat{A}^2} - \expval*{\hat{A}}^2$ is the
variance of the observable $\hat{A}$, and $\expval*{\hat{A}}$ is the
expectation value of $\hat{A}$. To achieve the minimum estimator variance
in $\hat{\theta}$, we should pick an observable $\hat{A}$ that has a
intrinsically low variance (numerator) and high sensitivity to changes in
$\theta$ (denominator). For separable states, the estimator variance $(\Delta
\theta)^2$ is bounded from below by the so-called standard quantum limit (SQL),
which for an $N$-qubit state, scales asymptotically as $\mathcal{O}(1/N)$. On
the other hand, for entangled states, the bound on $(\Delta \theta)^2$ scales asymptotically
as $\mathcal{O}(1/N^{2})$, which is the so-called Heisenberg limit
(HL)~\cite{Shettell2020}.

\noindent
Given a quantum state $\rho$ and a Hamiltonian $\hat{H}$ generating the
encoding unitary $U(\theta)$, when there are no restrictions on the
choice for $\hat{A}$, minimizing the expression in
Eq.~\ref{eq:estimator_variance_for_theta} over all possible quantum
measurement observables, i.e., positive operator-value measures
(POVMs), recovers the smallest achievable single-shot variance $(\Delta \theta)^2$ for $\rho$ and
$\hat{H}$, known as the quantum Cram\'{e}r-Rao bound (QCRB)~\cite{Helstrom1969,
Braunstein1994,Paris2009,Holevo2011}
\begin{align}
    (\Delta \theta)^2 \geq \frac{1}{F_{Q}(\rho, \hat{H})}.
    \label{eq:cramer_rao_bound}
\end{align}
\noindent
Here, $F_{Q}$ is the quantum Fisher information
(QFI)~\cite{Paris2009,Petz2011}, which quantifies the state's sensitivity to
small variations in $\theta$. In this study, we consider a Hamiltonian
$\hat{H}$ that takes form
\begin{align}
    \hat{H} = (\hat{X}_{1} \otimes \mathds{1}_{2} + \mathds{1}_{1} \otimes \hat{Z}_{2})/2,
\end{align}
\noindent
which is a locally rotated version of $\hat{H}$ used for the Bell state
$\ket{\Phi^{(+)}}$, as $\ket{\Gamma_{\pi}} = (H \otimes
\mathds{1})\ket*{\mathrm{\Phi^{(+)}}}$~\cite{Toth2014}. The unitary
transformation then becomes:
\begin{align}
    U(\theta) &= e^{-i\hat{H}\theta/\hbar} =  R^{(1)}_x(\theta) \otimes R^{(2)}_z(\theta).
    \label{eq:rx_rz}
\end{align}
\noindent
In our experiment, this unitary is realized through a configuration of a
QWP-HWP-QWP chain; one set for $R_x(\theta)$ and another for $R_z(\theta)$ (see
Fig.~\ref{fig:experiment_schematic}). By changing the waveplate angles we can
target an arbitrary value for the phase $\theta$.
Appendix~\ref{sec:single_q_rotation_gates_with_pol_optics} gives the relation
between $\theta$ and the waveplate angles. Practically, we should operate at a
point where the observable's expectation value has the steepest gradient,
i.e., highest sensitivity to changes in $\theta$. Theoretically, the QFI
for the 2-qubit weighted graph state, encoded via the unitary operator in
Eq.~\ref{eq:rx_rz} depends only on the graph weight $\phi_{12}$ and is
independent of the encoded phase $\theta$ (see
Appendix~\ref{sec:closed_form_qfi} for the derivation). We fix the sensing
phase at $\theta=0$, as small $\theta$ represents the most interesting regimes
for quantum metrology and its advantage over classical methods. This choice
does not limit generality, as any arbitrary phase $\theta$ can be shifted to
this operating point via an offset phase and we are ultimately interested not
in $\theta$ but in the precision $(\Delta \theta)^2$ around $\theta$.

\subsection{Estimator variance for local Pauli measurements}
To experimentally evaluate the metrological performance of our weighted graph
states $\ket*{\Gamma_{\phi_{12}}}$, we experimentally determine the estimator
variance $(\Delta \theta)^2$ in Eq.~\ref{eq:estimator_variance_for_theta} by
measuring a set of observable operators $\hat{A}$. We test two distinct sets of
local measurements, progressively relaxing the constraints dictated by our
experimental setup. Initially, we restrict the observable $\hat{A}$ to be a
tensor product of single-qubit Pauli operators $\hat{A} = \hat{A}_{1} \otimes
\hat{A}_{2}$ , where $\hat{A}_i \in \{\mathds{1}, \hat{X}, \hat{Y}, \hat{Z}\}$.
The corresponding measurement bases for these are products of the three
standard single-photon bases: rectilinear basis $\{\ket*{H}, \ket*{V}\}$,
diagonal basis $\{\ket*{D},\ket*{A}\}$ and circular basis $\{\ket*{R},
\ket*{L}\}$. We perform an exhaustive numerical search over all $4^2=16$
possible local Pauli product operators at each graph weight $\phi_{12}$ to find
the operator $\hat{A}$ that yields the smallest value for $(\Delta \theta)^2$.
In cases where multiple operators achieve the same minimum variance, we select
the one with the largest gradient magnitude
$\abs*{\partial_{\theta}\expval*{\hat{A}}}$, as this is generally preferred for
robust experimental implementation. More information about the optimal local
Pauli operators at the graph weights of interest, when $\theta=0$, can be found
in Appendix~\ref{sec:local_pauli_measurements}.

\begin{figure}[H]
    \centering
    \begin{subfigure}[t]{\linewidth}
	\caption{\label{fig:local_paulis_observable_variance}}
	\includegraphics[width=\linewidth]{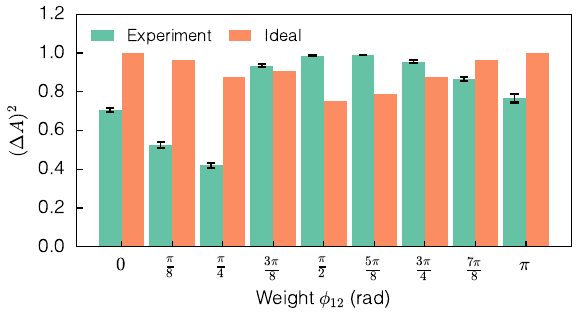}
    \end{subfigure}
    \begin{subfigure}[t]{\linewidth}
	\caption{\label{fig:local_paulis_observable_derivative}}
	\includegraphics[width=\linewidth]{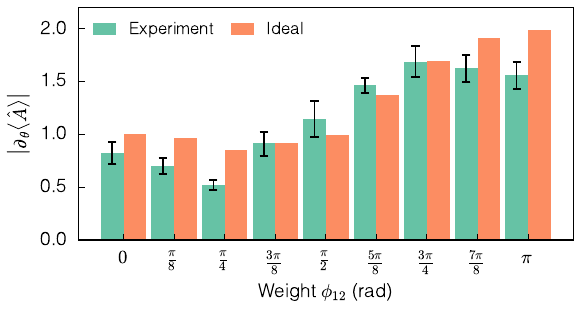}
    \end{subfigure}
    \caption{
	\textbf{(a)} Single-shot observable variance $(\Delta A)^2$, and
	\textbf{(b)} magnitude of the observable derivative
	$\abs*{\partial_{\theta}\expval*{\hat{A}}}$, measured via a 2-point
	finite difference method for the optimal local Pauli measurements as a
	function of the graph weight $\phi_{12}$ for the sensing phase $\theta=0$.
	The errors represent 95\% confidence intervals for the respective
	quantities.
	\label{fig:local_paulis_observable_variance_and_derivative}
    }
\end{figure}

To determine the estimator variance for each graph weight $\phi_{12}$ of
interest given an optimal observable operator $\hat{A}$, we begin by evaluating
the single-shot variance $(\Delta A)^2 = 1 - \expval*{\hat{A}}^2$ at
$\theta=0$. This is done by first collecting photon coincidence counts
$N_{\square\square}$ for each of the four projective measurements of $\hat{A}$,
i.e., $N_{++}$, $N_{+-}$, $N_{-+}$, and $N_{--}$. Each of the photon counts are
collected over a 10-second interval, and we build up data sets of six samples
for each projective measurement. The single-shot variance $(\Delta A)^2 = 1 -
\expval*{\hat{A}}^2$ is then calculated via a non-parametric bootstrap
resampling method~\cite{Efron1993} on the data set. More information about the
bootstrapping method can be found in Appendix~\ref{sec:bootstrapping}. The
values of $(\Delta A)^2$ for a range of weights are shown in
Fig.~\ref{fig:local_paulis_observable_variance}, where there is a degree of
correlation between the experiment and the ideal case.

\newpage
\noindent
It is important to note that $(\Delta A)^2$ is the noise originating from
quantum fluctuations only for a single pair of entangled photons in the state
$\ket*{\Gamma_{\phi_{12}}}$. However, in our experiment we are
measuring coincidence counts corresponding to many pairs of entangled photons
over a fixed period of 10 seconds, where the number of shots (photon pairs)
fluctuates due to Poisson noise from the SPDC process. This results in
additional noise in the data. In Appendix~\ref{sec:bootstrapping} we show that
for large mean coincidence counts, the impact of Poisson noise on the measured
value of $(\Delta A)^2$ is negligible, giving the value that can be expected if
one were to do sensing shot-by-shot. One way to do this is to detect both
outputs of the PBSs in the experiment and reduce the integration time from 10
secs to approximately the inverse of the count rate.

Each state's sensitivity $\abs*{\partial_{\theta}\expval*{\hat{A}}}$ is also
measured around the operating point $\theta=0$. The exact derivative for our
phase encoding unitary in Eq.~\ref{eq:rx_rz} can be evaluated via the
parameter-shift rule~\cite{Crooks2019}. However, to reduce the number of
required measurements, we approximate the derivative using a 2-point finite
difference method with a fixed shift $h=5^{\circ}$.
\begin{align}
    \eval{\partial_{\theta}\expval*{\hat{A}}}_{\theta=\theta^{\ast}} \approx \frac{\expval*{\hat{A}}(\theta^{\ast} + h) - \expval*{\hat{A}}(\theta^{\ast} - h)}{2h}, 
    \label{eq:2_point_finite_difference}
\end{align}
\noindent
where the argument $(\theta \pm h)$ implies the rotation $U(\theta \pm h) =
R^{(1)}_{x}(\theta \pm h) \otimes R^{(2)}_{z}(\theta \pm h)$. This approach
requires only two experimental evaluations of $\ev*{\hat{A}}$ around the point
of interest $\theta^{\ast}$, and has truncation error of order
$\mathcal{O}(h^2)$. This is a pragmatic choice, as the parameter-shift rule
would require four evaluations of $\expval*{\hat{A}}$, which requires more
stringent conditions on the stability of $\varphi^{\prime}$ in our experiment.

\begin{figure}[H]
    \includegraphics[width=\linewidth]{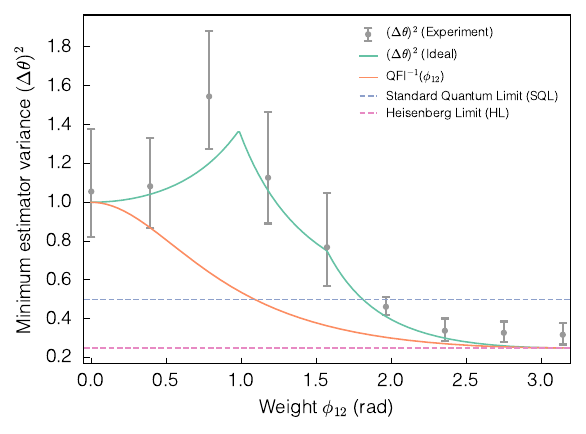}
    \caption{
	The estimator variance $(\Delta \theta)^2$ for the optimal local Pauli
	measurements. The grey dashed horizontal line represents the standard
	quantum limit (SQL), where $(\Delta \theta)^2=1/2$. The pink dashed
	horizontal line represents the Heisenberg limit (HL), $(\Delta
	\theta)^2 = 1/4$. The errors represent 95\% confidence intervals of the
	bootstrap samples for the respective quantities.
    }
    \label{fig:local_paulis_estimator_variance}
\end{figure}

Similar to the observable variance, the observable derivative in
Eq.~\ref{eq:2_point_finite_difference} is calculated with a non-parametric
bootstrap resampling method from the 6-sample data sets of the projective
measurements for $\expval*{\hat{A}}$ at $\theta=-5^{\circ}$ and
$\theta=5^{\circ}$. These results are shown
Fig.~\ref{fig:local_paulis_observable_derivative} and show reasonable
agreement between the experiment and ideal case. Further details about the
optimal local Pauli operators for each graph weight can be found in
Appendix~\ref{sec:local_pauli_measurements}. Finally, to evaluate the estimator
variance $(\Delta \theta)^2$ we combine the three data sets for
$\expval*{\hat{A}}$, at $\theta=\{-5^{\circ},\>0^{\circ},\>5^{\circ}\}$ and
bootstrap the data sets to calculate the estimator variance $(\Delta \theta)^2$
in Eq.~\ref{eq:estimator_variance_for_theta}. The results of this are shown in
Fig.~\ref{fig:local_paulis_estimator_variance}. We observe a close quantitative
agreement between the experimental results and theoretical predictions. For the
graph weights $\phi_{12}=\pi, \>7\pi/8,\>3\pi/4,\text{ and }\>5\pi/8$, the
experimental estimator variance $(\Delta \theta)^2$ is well below the SQL,
highlighting a gain in precision even for non-optimal graph weights $\pi >
\phi_{12} \geq 5\pi/8$, which correspond to non-maximally-entangled states. The
data point for $\phi_{12}=\pi/4$ is an outlier, potentially due to the
observable derivative being smallest in magnitude $\sim 0.85$ (theory), thus
the most susceptible to experimental errors.

\clearpage

\vspace{-3ex}

\subsection{Estimator variance for local general-axis measurements}
We now relax the constraint on the measurements, allowing $\hat{A}$ to be any general-axis local
operator, corresponding to a projective measurement along an arbitrary
direction $(\beta_{i}, \alpha_{i})$ on the Bloch sphere for each qubit ($i=1,2$). The
observable $\hat{A}$ takes the form:
\begin{align}
    &\hat{A} = \hat{A}_{1}(\beta_{1}, \alpha_{1}) \otimes \hat{A}_{2}(\beta_{2}, \alpha_{2}), \nonumber \\
    &\hat{A}_{i}(\beta_{i}, \alpha_{i}) := \dyad{\beta_{i}, \alpha_{i}, +} -  \dyad{\beta_{i}, \alpha_{i}, - }.
\end{align}
\noindent
Here, the projectors are formed from the general-axis basis states, given by
\begin{align}
    \ket*{\beta, \alpha, +} &:=\cos(\beta/2)\ket*{0} + \sin(\beta/2)e^{i \alpha}\ket*{1}, \nonumber \\ 
    \ket*{\beta, \alpha, -} &:=-\sin(\beta/2)\ket*{0} + \cos(\beta/2)e^{i \alpha}\ket*{1},
    \label{eq:general_pauli_basis_states}
\end{align}
\noindent
where $\beta$ and $\alpha$ are the polar and azimuthal angles, respectively. We
use this parameterization to perform numerical optimization for each graph
weight $\phi_{12}$ to find the optimal set of angles $\{\beta_{1}, \alpha_{1},
\beta_{2}, \alpha_{2}\}$ that minimize $(\Delta \theta)^2$. To guide the search
towards experimentally robust optima, a soft penalty is added to encourage
solutions with a large gradient magnitude. Experimentally, these general-axis
measurements can be realized by setting specific angles for the HWP and QWP acting
on each photon at the measurement stage. The wave plate angles are derived from
the optimal polar and azimuthal angles $\{(\beta_{i},\alpha_{i})\}$ by another
numerical optimization routine (see
Appendix~\ref{sec:general_axis_measurements} for more details). To
experimentally evaluate the estimator variance for each graph weight
$\phi_{12}$ of interest given an optimal general-axis measurement, we follow
the same procedure as in the local Pauli measurements; we collect data sets of
the projective measurements for $\expval*{\hat{A}}$, at
$\theta=\{-5^{\circ},0^{\circ},5^{\circ}\}$ and bootstrap the data sets to
calculate the observable variance, observable derivative, and  estimator
variance $(\Delta \theta)^2$ in Eq.~\ref{eq:estimator_variance_for_theta}.

\begin{figure}[H]
    \centering
    \begin{subfigure}[t]{\linewidth}
	\caption{\label{fig:general_axis_observable_variance}}
	\includegraphics[width=\linewidth]{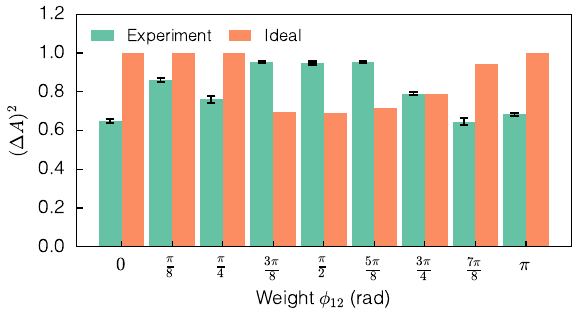}
    \end{subfigure}
    \begin{subfigure}[t]{\linewidth}
	\caption{\label{fig:general_axis_observable_derivative}}
	\includegraphics[width=\linewidth]{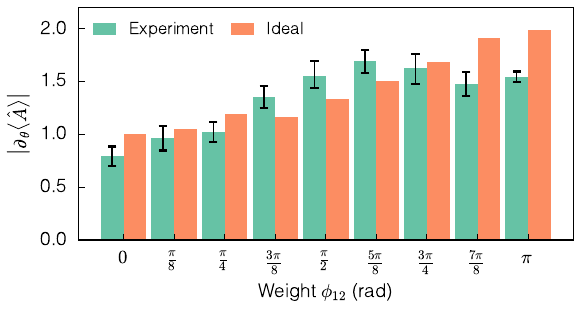}
    \end{subfigure}
    \caption{
	\textbf{(a)} Single-shot observable variance $(\Delta A)^2$, and
	\textbf{(b)} magnitude of the observable derivative
	$\abs*{\partial_{\theta}\expval*{\hat{A}}}$ measured via a 2-point
	finite difference method for the optimal local general-axis
	measurements as a function of the graph weight $\phi_{12}$ for the sensing
	phase $\theta=0$. The errors represent 95\% confidence intervals of the
	bootstrap samples for the respective quantities.
	\label{fig:general_axis_observable_variance_and_derivative}
    }
\end{figure}

\begin{figure}[H]
    \centering
    \includegraphics[width=\linewidth]{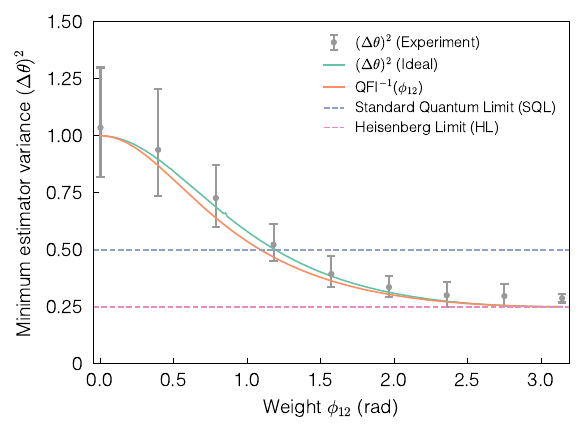}
    \caption{
	The estimator variance $(\Delta \theta)^2$ for the optimal local
	general-axis measurements. The grey dashed horizontal line represents
	the SQL, $(\Delta \theta)^2=1/2$. The pink dashed horizontal line
	represents the HL, $(\Delta \theta)^2 = 1/4$. The errors represent 95\%
	confidence intervals for the respective quantities.
    }
    \label{fig:general_axis_estimator_variance}
\end{figure}

\noindent
The comparison between the experimental and ideal observable variance $(\Delta
A)^2$ and sensitivity $\abs*{\partial_{\theta}\expval*{\hat{A}}}$ at $\theta=0$
for the optimal local general-axis operators for each graph weight $\phi_{12}$
are in shown in Fig.~\ref{fig:general_axis_observable_variance} and
Fig.~\ref{fig:general_axis_observable_derivative}, respectively. As with the
local Pauli measurements, one can see a reasonable agreement between the
experimental and ideal cases. Fig.~\ref{fig:general_axis_estimator_variance}
compares the experimental and ideal estimator variance $(\Delta \theta)^2$ for
local general-axis measurements. Once again, there is a close degree of
quantitative agreement between experimental results and theoretical
predictions. Using general-axis measurements has the effect of bringing the
estimator variance closer to the QCRB. Moreover, in contrast to the local Pauli
measurements, there is a larger range of graph weights that are below the SQL
($\pi \geq \phi_{12} \geq \pi/2$); the data point at $\phi_{12}=\pi/2$ has now
been pushed below the SQL.

\clearpage

\vspace{-3ex}

\subsection{Practical considerations}

The systematic mismatch between the experimentally measured values and the
theoretical predictions in
Fig.~\ref{fig:local_paulis_observable_variance_and_derivative} and
Fig.~\ref{fig:general_axis_observable_variance_and_derivative} can be explained
by an effective offset in the implemented graph weight and by small
miscalibrations in the waveplates used to realize the encoding unitary. In the
experiment, the weight $\phi_{12}$ is set by an interferometer and is therefore
sensitive to residual phases in each path and to slow drifts during data
acquisition; the reconstructed density matrices in
Fig.~\ref{fig:graph_state_s10} and their corresponding fidelities already
highlight this effect. Importantly, while a QWP-HWP-QWP configuration
implements the intended single-qubit operation up to an overall (global) phase,
any \emph{path-dependent} phase introduced by the optical elements appears as
an additional relative phase between interferometer arms and therefore shifts
the effective graph weight, i.e.\ $\phi_{12}\mapsto\phi_{12}^{\mathrm{eff}}$.
For example, the QWP-HWP-QWP configuration used to implement the identity
operation in path $r_2$ is equivalent to the identity only up to a constant
phase factor (here $e^{\pm i\pi/2}$); when this phase is not common to both
paths it contributes directly to the relative arm phase. In addition, the
effective rotation axes and zero-angle references of the QWP-HWP-QWP
configurations used to realize the identity operation for photon 2 in $r_2$,
$R_{z}(\varphi'_{12})$ for photon 2 in $l_2$, and the encoding rotations
$R_{x}(\theta)$ for photon 1 and $R_{z}(\theta)$ for photon 2 can be shifted by
small mounting and calibration offsets. Consequently, these systematic shifts
in the graph weight and in the realized encoding unitary manifest predominantly
as deviations in the measured observable derivative and variance.


\section{Conclusion\label{sec:conclusion}}
In summary, we have demonstrated a proof-of-concept scheme for realizing a
photonic two-qubit weighted graph state with an arbitrarily tunable weight,
using linear optics and a single polarization-entangled photon-pair source
based on SPDC. We successfully generated weighted graph states with
different weights with a fidelity in excess of $80\%$. Following the
theoretical framework in Ref.~\cite{Alexander2025}, establishing the robustness
of weighted graph states for metrological applications, we experimentally
verified that considerably less entanglement is required to achieve a quantum
advantage for metrology. In the first case, this was done by experimentally
verifying that under local Pauli measurements, weighted graph states with
$\pi \geq \phi_{12} \geq 5\pi/8$ are sufficient to yield a gain in precision beyond the
SQL. Relaxing the restriction and allowing local
measurements along an arbitrary axis led to a wider range of graph weights
$\pi \geq \phi_{12} \geq \pi/2$ that correspond to weighted graph states achieving a gain
in precision below the SQL.

Our results highlight the practical utility of weighted graph states by
demonstrating that near-maximal entanglement is not a prerequisite for
achieving quantum-enhanced sensing. This is important for real-world
applications where noise and decoherence often limit the generation and
maintenance of highly entangled states. By showing that moderate levels of
entanglement ($\phi_{12} \geq \pi/2$) are sufficient to surpass the classical
limit, our work reflects a resource-efficient route to quantum metrology. The
experimental results presented here were achieved with weighted graph states
with fidelities in excess of $80\%$. We anticipate that the fidelity of the
generated weighted graph states can be significantly improved via
standard methods, such as utilizing high-efficiency entangled-photon sources,
implementing spectral mode matching, and using walk-off compensators for
temporal matching of horizontally- and vertically-polarized light cones emitted
by the cascaded BBOs~\cite{Altepeter2005,Akselrod2007,Rangarajan2009}. These
improvements could potentially have direct downstream effects, resulting in an
improved measured estimator variance $(\Delta \theta)^2$.

Directions for future work include extending this generation scheme to larger
systems, building upon established methods for standard graph states such as
those presented in Ref.~\cite{Park2007,Xiong2011}. The use of moderately sized
weighted graph states could be immediately beneficial for quantum computing
paradigms like fusion-based quantum
computation~\cite{Bartolucci2023,Thomas2024}, as weighted states are already
known to increase the success probability for the fusion
process~\cite{Rimock2024}. Beyond scaling the resource state, the metrology
protocol itself could be enhanced in two primary ways. First, it could be extended
to multiparameter estimation~\cite{Proctor2018,Ge2018,Zhang2021,Kim2024}, where the encoding unitary possesses two or more
independent parameters instead of one, such as simultaneously sensing the $x$
and $y$ components of a magnetic or electric
field~\cite{Kaubruegger2023,Valeri2023,Le2023,Zhang2024}. Second, the protocol
could be automated for larger weighted graph states and/or multiple parameters
via an end-to-end variational framework for quantum sensing
protocols~\cite{Valeri2023,Le2023,MacLellan2024}.

\section*{Acknowledgments}
This research was supported by the South African National Research
Foundation, the South African Council for Scientific and Industrial Research,
and the South African Department of Science and Innovation through its Quantum
Initiative program (SAQuTI).

\bibliographystyle{apsrev4-2}
\bibliography{main.bib}

@article{James2001,
  title = {Measurement of qubits},
  author = {James, Daniel F. V. and Kwiat, Paul G. and Munro, William J. and White, Andrew G.},
  journal = {Physical Review A},
  volume = {64},
  issue = {5},
  pages = {052312},
  numpages = {15},
  year = {2001},
  month = {Oct},
  publisher = {American Physical Society},
  doi = {10.1103/PhysRevA.64.052312},
  url = {https://link.aps.org/doi/10.1103/PhysRevA.64.052312}
}

@article{Gao2010,
  title         = {Experimental demonstration of a hyper-entangled ten-qubit Schr\"{o}dinger cat state},
  volume        = {6},
  issn          = {1745-2481},
  url           = {http://dx.doi.org/10.1038/nphys1603},
  number        = {5},
  journal       = {Nature Physics},
  publisher     = {Springer Science and Business Media LLC},
  author        = {Gao, Wei-Bo and others},
  year          = {2010},
  month         = mar,
  pages         = {331–335}
}

@article{Li2020,
  author        = {Li, Jin-Peng and others},
  title         = {Multiphoton Graph States from a Solid-State Single-Photon Source},
  journal       = {ACS Photonics},
  volume        = {7},
  number        = {7},
  pages         = {1603--1610},
  year          = {2020},
  url           = {https://doi.org/10.1021/acsphotonics.0c00192}
}

@article{Park2007,
  author        = {Hee Su Park and Jaeyoon Cho and Jae Yong Lee and Dong-Hoon Lee and Sang-Kyung Choi},
  journal       = {Optics Express},
  number        = {26},
  pages         = {17960--17966},
  publisher     = {Optica Publishing Group},
  title         = {Two-photon four-qubit cluster state generation based on a polarization-entangled photon pair},
  volume        = {15},
  month         = {Dec},
  year          = {2007},
  url           = {https://opg.optica.org/oe/abstract.cfm?URI=oe-15-26-17960}
}

@article{Lee2012,
  author        = {Sang Min Lee and others},
  journal       = {Optics Express},
  number        = {7},
  pages         = {6915--6926},
  publisher     = {Optica Publishing Group},
  title         = {Experimental realization of a four-photon seven-qubit graph state for one-way quantum computation},
  volume        = {20},
  month         = {Mar},
  year          = {2012},
  url           = {https://opg.optica.org/oe/abstract.cfm?URI=oe-20-7-6915}
}

@article{Zhong2018,
  title         = {12-Photon Entanglement and Scalable Scattershot Boson Sampling with Optimal Entangled-Photon Pairs from Parametric Down-Conversion},
  author        = {Zhong, Han-Sen and others},
  journal       = {Physical Review Letters},
  volume        = {121},
  issue         = {25},
  pages         = {250505},
  numpages      = {6},
  year          = {2018},
  month         = {Dec},
  publisher     = {American Physical Society},
  url           = {https://link.aps.org/doi/10.1103/PhysRevLett.121.250505}
}

@article{Wang2025,
  author={Wang, Ze and others},
  title={Large-scale cluster quantum microcombs},
  journal={Light: Science {\&} Applications},
  year={2025},
  month={Apr},
  day={16},
  volume={14},
  number={1},
  pages={164},
  issn={2047-7538},
  doi={10.1038/s41377-025-01812-2},
  url={https://doi.org/10.1038/s41377-025-01812-2}
}

@article{Wang2018,
  title         = {18-Qubit Entanglement with Six Photons' Three Degrees of Freedom},
  author        = {Wang, Xi-Lin and others},
  journal       = {Physical Review Letters},
  volume        = {120},
  issue         = {26},
  pages         = {260502},
  numpages      = {5},
  year          = {2018},
  month         = {Jun},
  publisher     = {American Physical Society},
  url           = {https://link.aps.org/doi/10.1103/PhysRevLett.120.260502}
}

@article{Pont2024,
author={Pont, Mathias and others},
  title={High-fidelity four-photon GHZ states on chip},
  journal={npj Quantum Information},
  year={2024},
  month={May},
  day={15},
  volume={10},
  number={1},
  pages={50},
  issn={2056-6387},
  doi={10.1038/s41534-024-00830-z},
  url={https://doi.org/10.1038/s41534-024-00830-z}
}

@article{Yang2025,
  author={Yang, Yang and others},
  title={Programmable quantum circuits in a large-scale photonic waveguide array},
  journal={npj Quantum Information},
  year={2025},
  month={Feb},
  day={03},
  volume={11},
  number={1},
  pages={19},
  issn={2056-6387},
  doi={10.1038/s41534-024-00934-6},
  url={https://doi.org/10.1038/s41534-024-00934-6}
}

@article{Bao2023,
  author        = {Bao, Jueming and others},
  title         = {Very-large-scale integrated quantum graph photonics},
  journal       = {Nature Photonics},
  year          = {2023},
  month         = {Jul},
  day           = {01},
  volume        = {17},
  number        = {7},
  pages         = {573--581},
  issn          = {1749-4893},
  url           = {https://doi.org/10.1038/s41566-023-01187-z}
}

@article{Zhong2020,
  author        = {Han-Sen Zhong and others},
  title         = {Quantum computational advantage using photons},
  journal       = {Science},
  volume        = {370},
  number        = {6523},
  pages         = {1460--1463},
  year          = {2020},
  url           = {https://www.science.org/doi/abs/10.1126/science.abe8770}
}

@article{Obrien2007,
  author        = {Jeremy L. O'Brien},
  title         = {Optical Quantum Computing},
  journal       = {Science},
  volume        = {318},
  number        = {5856},
  pages         = {1567--1570},
  year          = {2007},
  url           = {https://www.science.org/doi/abs/10.1126/science.1142892}
}

@article{Wang2019,
  title         = {Integrated photonic quantum technologies},
  volume        = {14},
  issn          = {1749-4893},
  url           = {http://dx.doi.org/10.1038/s41566-019-0532-1},
  number        = {5},
  journal       = {Nature Photonics},
  publisher     = {Springer Science and Business Media LLC},
  author        = {Wang, Jianwei and Sciarrino, Fabio and Laing, Anthony and Thompson, Mark G.},
  year          = {2019},
  month         = oct,
  pages         = {273–284}
}

@article{Lanyon2013,
  title = {Measurement-Based Quantum Computation with Trapped Ions},
  author = {Lanyon, B. P. and others},
  journal = {Physical Review Letters},
  volume = {111},
  issue = {21},
  pages = {210501},
  numpages = {5},
  year = {2013},
  month = {Nov},
  publisher = {American Physical Society},
  doi = {10.1103/PhysRevLett.111.210501},
  url = {https://link.aps.org/doi/10.1103/PhysRevLett.111.210501}
}

@article{Istrati2020,
  author={Istrati, D. and others},
  title={Sequential generation of linear cluster states from a single photon emitter},
  journal={Nature Communications},
  year={2020},
  month={Oct},
  day={30},
  volume={11},
  number={1},
  pages={5501},
  issn={2041-1723},
  doi={10.1038/s41467-020-19341-4},
  url={https://doi.org/10.1038/s41467-020-19341-4}
}

@article{Cogan2023,
  author        = {Cogan, Dan and Su, Zu-En and Kenneth, Oded and Gershoni, David},
  title         = {Deterministic generation of indistinguishable photons in a cluster state},
  journal       = {Nature Photonics},
  year          = {2023},
  month         = {Apr},
  day           = {01},
  volume        = {17},
  number        = {4},
  pages         = {324--329},
  issn          = {1749-4893},
  url           = {https://doi.org/10.1038/s41566-022-01152-2}
}

@article{Cramer2016,
  author={Cramer, J. and others},
  title={Repeated quantum error correction on a continuously encoded qubit by real-time feedback},
  journal={Nature Communications},
  year={2016},
  month={May},
  day={05},
  volume={7},
  number={1},
  pages={11526},
  issn={2041-1723},
  doi={10.1038/ncomms11526},
  url={https://doi.org/10.1038/ncomms11526}
}

@article{Mandel2003,
  title         = {Controlled collisions for multi-particle entanglement of optically trapped atoms},
  volume        = {425},
  issn          = {1476-4687},
  url           = {http://dx.doi.org/10.1038/nature02008},
  number        = {6961},
  journal       = {Nature},
  publisher     = {Springer Science and Business Media LLC},
  author        = {Mandel, Olaf and others},
  year          = {2003},
  month         = oct,
  pages         = {937–940}
}

@article{Thomas2024,
  title         = {Fusion of deterministically generated photonic graph states},
  volume        = {629},
  issn          = {1476-4687},
  url           = {http://dx.doi.org/10.1038/s41586-024-07357-5},
  number        = {8012},
  journal       = {Nature},
  publisher     = {Springer Science and Business Media LLC},
  author        = {Thomas, Philip and Ruscio, Leonardo and Morin, Olivier and Rempe, Gerhard},
  year          = {2024},
  month         = may,
  pages         = {567–572}
}

@article{Thomas2022,
  title         = {Efficient generation of entangled multiphoton graph states from a single atom},
  volume        = {608},
  issn          = {1476-4687},
  url           = {http://dx.doi.org/10.1038/s41586-022-04987-5},
  number        = {7924},
  journal       = {Nature},
  publisher     = {Springer Science and Business Media LLC},
  author        = {Thomas, Philip and Ruscio, Leonardo and Morin, Olivier and Rempe, Gerhard},
  year          = {2022},
  month         = aug,
  pages         = {677–681}
}

@article{Mooney2019,
  author={Mooney, Gary J. and Hill, Charles D. and Hollenberg, Lloyd C. L.},
  title={Entanglement in a 20-Qubit Superconducting Quantum Computer},
  journal={Scientific Reports},
  year={2019},
  month={Sep},
  day={17},
  volume={9},
  number={1},
  pages={13465},
  issn={2045-2322},
  doi={10.1038/s41598-019-49805-7},
  url={https://doi.org/10.1038/s41598-019-49805-7}
}

@article{Song2017,
  title = {10-Qubit Entanglement and Parallel Logic Operations with a Superconducting Circuit},
  author = {Song, Chao and others},
  journal = {Physical Review Letters},
  volume = {119},
  issue = {18},
  pages = {180511},
  numpages = {6},
  year = {2017},
  month = {Nov},
  publisher = {American Physical Society},
  doi = {10.1103/PhysRevLett.119.180511},
  url = {https://link.aps.org/doi/10.1103/PhysRevLett.119.180511}
}

@article{Plato2008,
  title         = {Random circuits by measurements on weighted graph states},
  author        = {Plato, A. Douglas K. and Dahlsten, Oscar C. and Plenio, Martin B.},
  journal       = {Physical Review A},
  volume        = {78},
  issue         = {4},
  pages         = {042332},
  numpages      = {7},
  year          = {2008},
  month         = {Oct},
  publisher     = {American Physical Society},
  url           = {https://link.aps.org/doi/10.1103/PhysRevA.78.042332}
}

@article{Xue2012,
  title         = {Spin-squeezing property of weighted graph states},
  author        = {Xue, Peng},
  journal       = {Physical Review A},
  volume        = {86},
  issue         = {2},
  pages         = {023812},
  numpages      = {8},
  year          = {2012},
  month         = {Aug},
  publisher     = {American Physical Society},
  url           = {https://link.aps.org/doi/10.1103/PhysRevA.86.023812}
}

@article{Hayashi2019,
  url           = {https://doi.org/10.1088/1367-2630/ab3d88},
  year          = {2019},
  month         = {sep},
  publisher     = {IOP Publishing},
  volume        = {21},
  number        = {9},
  pages         = {093060},
  author        = {Hayashi, Masahito and Takeuchi, Yuki},
  title         = {Verifying commuting quantum computations via fidelity estimation of weighted graph states},
  journal       = {New Journal of Physics}
}

@article{Ghost2024,
  title         = {Entanglement of weighted graphs uncovering transitions in variable-range interacting models},
  author        = {Ghosh, Debkanta and Das Agarwal, Keshav and Halder, Pritam and Sen(De), Aditi},
  journal       = {Physical Review A},
  volume        = {110},
  issue         = {2},
  pages         = {022431},
  numpages      = {13},
  year          = {2024},
  month         = {Aug},
  publisher     = {American Physical Society},
  url           = {https://link.aps.org/doi/10.1103/PhysRevA.110.022431}
}

@article{Tame2009,
  title = {Compact Toffoli gate using weighted graph states},
  author={Tame, M. S. and Özdemir, {\c{S}}. K. and Koashi, M. and Imoto, N. and Kim, M. S.},
  journal = {Physical Review A},
  volume = {79},
  issue = {2},
  pages = {020302},
  numpages = {4},
  year = {2009},
  month = {Feb},
  publisher = {American Physical Society},
  doi = {10.1103/PhysRevA.79.020302},
  url = {https://link.aps.org/doi/10.1103/PhysRevA.79.020302}
}

@article{Anders2006,
  title         = {Ground-State Approximation for Strongly Interacting Spin Systems in Arbitrary Spatial Dimension},
  author        = {Anders, S. and Plenio, M. B. and D\"ur, W. and Verstraete, F. and Briegel, H.-J.},
  journal       = {Physical Review Letters},
  volume        = {97},
  issue         = {10},
  pages         = {107206},
  numpages      = {4},
  year          = {2006},
  month         = {Sep},
  publisher     = {American Physical Society},
  url           = {https://link.aps.org/doi/10.1103/PhysRevLett.97.107206}
}

@article{Anders2007,
  url           = {https://doi.org/10.1088/1367-2630/9/10/361},
  year          = {2007},
  month         = {oct},
  publisher     = {},
  volume        = {9},
  number        = {10},
  pages         = {361},
  author        = {Anders, Simon and Briegel, Hans J and D\"{u}r, Wolfgang},
  title         = {A variational method based on weighted graph states},
  journal       = {New Journal of Physics}
}

@article{Firstenberg2013,
  author={Firstenberg, Ofer and others},
  title={Attractive photons in a quantum nonlinear medium},
  journal={Nature},
  year={2013},
  month={Oct},
  day={01},
  volume={502},
  number={7469},
  pages={71-75},
  issn={1476-4687},
  doi={10.1038/nature12512},
  url={https://doi.org/10.1038/nature12512}
}

@article{Firstenberg2016,
  url           = {https://doi.org/10.1088/0953-4075/49/15/152003},
  year          = {2016},
  month         = {jun},
  publisher     = {IOP Publishing},
  volume        = {49},
  number        = {15},
  pages         = {152003},
  author        = {Firstenberg, O and Adams, C S and Hofferberth, S},
  title         = {Nonlinear quantum optics mediated by Rydberg interactions},
  journal       = {Journal of Physics B: Atomic, Molecular and Optical Physics}
}

@article{Thompson2017,
  title         = {Symmetry-protected collisions between strongly interacting photons},
  author        = {Thompson, Jeff D and others},
  number        = {7640},
  volume        = {542},
  month         = {February},
  year          = {2017},
  journal       = {Nature},
  issn          = {0028-0836},
  pages         = {206--209},
  url           = {https://doi.org/10.1038/nature20823}
}

@article{Tiarks2018,
  title         = {A photon–photon quantum gate based on Rydberg interactions},
  volume        = {15},
  issn          = {1745-2481},
  url           = {http://dx.doi.org/10.1038/s41567-018-0313-7},
  number        = {2},
  journal       = {Nature Physics},
  publisher     = {Springer Science and Business Media LLC},
  author        = {Tiarks, Daniel and Schmidt-Eberle, Steffen and Stolz, Thomas and Rempe, Gerhard and D\"{u}rr, Stephan},
  year          = {2018},
  month         = oct,
  pages         = {124–126}
}

@article{Sagona2020,
  title         = {Conditional $\ensuremath{\pi}$-Phase Shift of Single-Photon-Level Pulses at Room Temperature},
  author        = {Sagona-Stophel, Steven and Shahrokhshahi, Reihaneh and Jordaan, Bertus and Namazi, Mehdi and Figueroa, Eden},
  journal       = {Physical Review Letters},
  volume        = {125},
  issue         = {24},
  pages         = {243601},
  numpages      = {6},
  year          = {2020},
  month         = {Dec},
  publisher     = {American Physical Society},
  url           = {https://link.aps.org/doi/10.1103/PhysRevLett.125.243601}
}

@article{Tiarks2016,
  author        = {Daniel Tiarks  and Steffen Schmidt  and Gerhard Rempe  and Stephan D\"{u}rr},
  title         = {Optical \ensuremath{\pi} phase shift created with a single-photon pulse},
  journal       = {Science Advances},
  volume        = {2},
  number        = {4},
  pages         = {e1600036},
  year          = {2016},
  url           = {https://www.science.org/doi/abs/10.1126/sciadv.1600036}
}

@article{Bartolucci2023,
  author        = {Bartolucci, Sara and others},
  title         = {Fusion-based quantum computation},
  journal       = {Nature Communications},
  year          = {2023},
  month         = {Feb},
  day           = {17},
  volume        = {14},
  number        = {1},
  pages         = {912},
  issn          = {2041-1723},
  url           = {https://doi.org/10.1038/s41467-023-36493-1}
}

@article{Dur2003,
  title         = {Multiparticle Entanglement Purification for Graph States},
  author        = {D\"ur, W. and Aschauer, H. and Briegel, H.-J.},
  journal       = {Physical Review Letters},
  volume        = {91},
  issue         = {10},
  pages         = {107903},
  numpages      = {4},
  year          = {2003},
  month         = {Sep},
  publisher     = {American Physical Society},
  url           = {https://link.aps.org/doi/10.1103/PhysRevLett.91.107903}
}

@article{Markham2008,
  title         = {Graph states for quantum secret sharing},
  author        = {Markham, Damian and Sanders, Barry C.},
  journal       = {Physical Review A},
  volume        = {78},
  issue         = {4},
  pages         = {042309},
  numpages      = {17},
  year          = {2008},
  month         = {Oct},
  publisher     = {American Physical Society},
  url           = {https://link.aps.org/doi/10.1103/PhysRevA.78.042309}
}

@article{Frantzeskakis2023,
  title         = {Extracting perfect GHZ states from imperfect weighted graph states via entanglement concentration},
  author        = {Frantzeskakis, Rafail and Liu, Chenxu and Raissi, Zahra and Barnes, Edwin and Economou, Sophia E.},
  journal       = {Physical Review Research},
  volume        = {5},
  issue         = {2},
  pages         = {023124},
  numpages      = {11},
  year          = {2023},
  month         = {May},
  publisher     = {American Physical Society},
  url           = {https://link.aps.org/doi/10.1103/PhysRevResearch.5.023124}
}

@article{Choudhury2013,
  author        = {Choudhury, Binayak S. and Dhara, Arpan},
  title         = {An entanglement concentration protocol for cluster states},
  journal       = {Quantum Information Processing},
  year          = {2013},
  month         = {Jul},
  day           = {01},
  volume        = {12},
  number        = {7},
  pages         = {2577--2585},
  issn          = {1573-1332},
  url           = {https://doi.org/10.1007/s11128-013-0549-1}
}

@article{Deng2012,
  title         = {Optimal nonlocal multipartite entanglement concentration based on projection measurements},
  author        = {Deng, Fu-Guo},
  journal       = {Physical Review A},
  volume        = {85},
  issue         = {2},
  pages         = {022311},
  numpages      = {5},
  year          = {2012},
  month         = {Feb},
  publisher     = {American Physical Society},
  url           = {https://link.aps.org/doi/10.1103/PhysRevA.85.022311}
}

@article{Zhou2012,
  title         = {Efficient entanglement concentration for arbitrary single-photon multimode W state},
  volume        = {30},
  issn          = {1520-8540},
  url           = {http://dx.doi.org/10.1364/JOSAB.30.000071},
  number        = {1},
  journal       = {Journal of the Optical Society of America B},
  publisher     = {Optica Publishing Group},
  author        = {Zhou, Lan and Sheng, Yu-Bo and Cheng, Wei-Wen and Gong, Long-Yan and Zhao, Sheng-Mei},
  year          = {2012},
  month         = dec,
  pages         = {71}
}

@article{Xiong2011,
  author        = {Wei Xiong and Liu Ye},
  journal       = {Journal of the Optical Society of America B},
  number        = {8},
  pages         = {2030--2037},
  publisher     = {Optica Publishing Group},
  title         = {Schemes for entanglement concentration of two unknown partially entangled states with cross-Kerr nonlinearity},
  volume        = {28},
  month         = {Aug},
  year          = {2011},
  url           = {https://opg.optica.org/josab/abstract.cfm?URI=josab-28-8-2030}
}

@article{Hwang2007,
  title         = {Practical scheme for non-postselection entanglement concentration using linear optical elements},
  journal       = {Physics Letters A},
  volume        = {369},
  number        = {4},
  pages         = {280--284},
  year          = {2007},
  issn          = {0375-9601},
  url           = {https://www.sciencedirect.com/science/article/pii/S0375960107006688},
  author        = {Myung-Joong Hwang and Yoon-Ho Kim}
}

@article{Paunkovi2002,
  title         = {Entanglement Concentration Using Quantum Statistics},
  author        = {Paunkovi\ifmmode \acute{c}\else \'{c}\fi{}, N. and Omar, Y. and Bose, S. and Vedral, V.},
  journal       = {Physical Review Letters},
  volume        = {88},
  issue         = {18},
  pages         = {187903},
  numpages      = {4},
  year          = {2002},
  month         = {Apr},
  publisher     = {American Physical Society},
  url           = {https://link.aps.org/doi/10.1103/PhysRevLett.88.187903}
}

@article{Sheng2012,
  title         = {Efficient single-photon-assisted entanglement concentration for partially entangled photon pairs},
  author        = {Sheng, Yu-Bo and Zhou, Lan and Zhao, Sheng-Mei and Zheng, Bao-Yu},
  journal       = {Physical Review A},
  volume        = {85},
  issue         = {1},
  pages         = {012307},
  numpages      = {7},
  year          = {2012},
  month         = {Jan},
  publisher     = {American Physical Society},
  url           = {https://link.aps.org/doi/10.1103/PhysRevA.85.012307}
}

@article{Zhao2003,
  title         = {Experimental Realization of Entanglement Concentration and a Quantum Repeater},
  author        = {Zhao, Zhi and Yang, Tao and Chen, Yu-Ao and Zhang, An-Ning and Pan, Jian-Wei},
  journal       = {Physical Review Letters},
  volume        = {90},
  issue         = {20},
  pages         = {207901},
  numpages      = {4},
  year          = {2003},
  month         = {May},
  publisher     = {American Physical Society},
  url           = {https://link.aps.org/doi/10.1103/PhysRevLett.90.207901}
}

@article{Shettell2020,
  title = {Graph States as a Resource for Quantum Metrology},
  author = {Shettell, Nathan and Markham, Damian},
  journal = {Physical Review Letters},
  volume = {124},
  issue = {11},
  pages = {110502},
  numpages = {6},
  year = {2020},
  month = {Mar},
  publisher = {American Physical Society},
  doi = {10.1103/PhysRevLett.124.110502},
  url = {https://link.aps.org/doi/10.1103/PhysRevLett.124.110502}
}

@article{Friis2017,
  title         = {Flexible resources for quantum metrology},
  volume        = {19},
  issn          = {1367-2630},
  url           = {http://dx.doi.org/10.1088/1367-2630/aa7144},
  number        = {6},
  journal       = {New Journal of Physics},
  publisher     = {IOP Publishing},
  author        = {Friis, Nicolai and others},
  year          = {2017},
  month         = jun,
  pages         = {063044}
}

@article{Walther2005,
  title         = {Experimental one-way quantum computing},
  volume        = {434},
  issn          = {1476-4687},
  url           = {http://dx.doi.org/10.1038/nature03347},
  number        = {7030},
  journal       = {Nature},
  publisher     = {Springer Science and Business Media LLC},
  author        = {Walther, P. and others},
  year          = {2005},
  month         = mar,
  pages         = {169–176}
}

@article{Briegel2001,
  title         = {Persistent Entanglement in Arrays of Interacting Particles},
  volume        = {86},
  issn          = {1079-7114},
  url           = {http://dx.doi.org/10.1103/PhysRevLett.86.910},
  number        = {5},
  journal       = {Physical Review Letters},
  publisher     = {American Physical Society (APS)},
  author        = {Briegel, Hans J. and Raussendorf, Robert},
  year          = {2001},
  month         = jan,
  pages         = {910–913}
}

@article{Raussendorf2003,
  title = {Measurement-based quantum computation on cluster states},
  author = {Raussendorf, Robert and Browne, Daniel E. and Briegel, Hans J.},
  journal = {Physical Review A},
  volume = {68},
  issue = {2},
  pages = {022312},
  numpages = {32},
  year = {2003},
  month = {Aug},
  publisher = {American Physical Society},
  doi = {10.1103/PhysRevA.68.022312},
  url = {https://link.aps.org/doi/10.1103/PhysRevA.68.022312}
}

@article{Raussendorf2001,
  title         = {A One-Way Quantum Computer},
  author        = {Raussendorf, Robert and Briegel, Hans J.},
  journal       = {Physical Review Letters},
  volume        = {86},
  issue         = {22},
  pages         = {5188--5191},
  numpages      = {0},
  year          = {2001},
  month         = {May},
  publisher     = {American Physical Society},
  url           = {https://link.aps.org/doi/10.1103/PhysRevLett.86.5188}
}

@article{Menicucci2006,
  title = {Universal Quantum Computation with Continuous-Variable Cluster States},
  author = {Menicucci, Nicolas C. and others},
  journal = {Physical Review Letters},
  volume = {97},
  issue = {11},
  pages = {110501},
  numpages = {4},
  year = {2006},
  month = {Sep},
  publisher = {American Physical Society},
  doi = {10.1103/PhysRevLett.97.110501},
  url = {https://link.aps.org/doi/10.1103/PhysRevLett.97.110501}
}

@article{Hartmann2007,
  url           = {https://doi.org/10.1088/0953-4075/40/9/S01},
  year          = {2007},
  month         = {apr},
  publisher     = {},
  volume        = {40},
  number        = {9},
  pages         = {S1},
  author        = {Hartmann, L and Calsamiglia, J and D\"{u}r, W and Briegel, H J},
  title         = {Weighted graph states and applications to spin chains, lattices and gases},
  journal       = {Journal of Physics B: Atomic, Molecular and Optical Physics}
}

@article{Dur2005,
  title = {Entanglement in Spin Chains and Lattices with Long-Range Ising-Type Interactions},
  author = {D\"ur, W. and Hartmann, L. and Hein, M. and Lewenstein, M. and Briegel, H.-J.},
  journal = {Physical Review Letters},
  volume = {94},
  issue = {9},
  pages = {097203},
  numpages = {4},
  year = {2005},
  month = {Mar},
  publisher = {American Physical Society},
  doi = {10.1103/PhysRevLett.94.097203},
  url = {https://link.aps.org/doi/10.1103/PhysRevLett.94.097203}
}

@article{Li2014,
  url           = {https://doi.org/10.1088/1612-2011/11/12/125201},
  year          = {2014},
  month         = {oct},
  publisher     = {IOP Publishing},
  volume        = {11},
  number        = {12},
  pages         = {125201},
  author        = {Li, Xihan and Ghose, Shohini},
  title         = {Hyperconcentration for multipartite entanglement via linear optics},
  journal       = {Laser Physics Letters}
}

@article{Simon1990,
  title         = {Minimal three-component SU(2) gadget for polarization optics},
  journal       = {Physics Letters A},
  volume        = {143},
  number        = {4},
  pages         = {165--169},
  year          = {1990},
  issn          = {0375-9601},
  url           = {https://www.sciencedirect.com/science/article/pii/0375960190907324},
  author        = {R. Simon and N. Mukunda}
}

@article{Reddy2014,
  author        = {Reddy, Salla Gangi and others},
  journal       = {Journal of the Optical Society of America A},
  number        = {3},
  pages         = {610--615},
  publisher     = {Optica Publishing Group},
  title         = {Measuring the Mueller matrix of an arbitrary optical element with a universal SU(2) polarization gadget},
  volume        = {31},
  month         = {Mar},
  year          = {2014},
  url           = {https://opg.optica.org/josaa/abstract.cfm?URI=josaa-31-3-610}
}

@article{Branning2009,
  author        = {Branning, D. and Bhandari, S. and Beck, M.},
  title         = {Low-cost coincidence-counting electronics for undergraduate quantum optics},
  journal       = {American Journal of Physics},
  volume        = {77},
  number        = {7},
  pages         = {667--670},
  year          = {2009},
  month         = {07},
  issn          = {0002-9505},
  url           = {https://doi.org/10.1119/1.3116803}
}

@article{Horodecki2009,
  title         = {Quantum entanglement},
  author        = {Horodecki, Ryszard and Horodecki, Pawe\l{} and Horodecki, Micha\l{} and Horodecki, Karol},
  journal       = {Review of Modern Physics},
  volume        = {81},
  issue         = {2},
  pages         = {865--942},
  numpages      = {0},
  year          = {2009},
  month         = {Jun},
  publisher     = {American Physical Society},
  url           = {https://link.aps.org/doi/10.1103/RevModPhys.81.865}
}

@article{Hill1997,
  title         = {Entanglement of a Pair of Quantum Bits},
  author        = {Hill, Sam A. and Wootters, William K.},
  journal       = {Physical Review Letters},
  volume        = {78},
  issue         = {26},
  pages         = {5022--5025},
  numpages      = {0},
  year          = {1997},
  month         = {Jun},
  publisher     = {American Physical Society},
  url           = {https://link.aps.org/doi/10.1103/PhysRevLett.78.5022}
}

@article{Romero2007,
  title         = {Direct measurement of concurrence for atomic two-qubit pure states},
  author        = {Romero, G. and L\'opez, C. E. and Lastra, F. and Solano, E. and Retamal, J. C.},
  journal       = {Physical Review A},
  volume        = {75},
  issue         = {3},
  pages         = {032303},
  numpages      = {4},
  year          = {2007},
  month         = {Mar},
  publisher     = {American Physical Society},
  url           = {https://link.aps.org/doi/10.1103/PhysRevA.75.032303}
}

@article{Paris2009,
  author        = {Paris, Matteo G. A.},
  title         = {QUANTUM ESTIMATION FOR QUANTUM TECHNOLOGY},
  journal       = {International Journal of Quantum Information},
  volume        = {07},
  pages         = {125--137},
  year          = {2009},
  url           = {https://doi.org/10.1142/S0219749909004839}
}

@article{Helstrom1969,
  author        = {{Helstrom}, Carl W.},
  title         = "{Quantum detection and estimation theory}",
  journal       = {Journal of Statistical Physics},
  year          = 1969,
  month         = jun,
  volume        = {1},
  number        = {2},
  pages         = {231--252}
}

@article{Braunstein1994,
  title         = {Statistical distance and the geometry of quantum states},
  author        = {Braunstein, Samuel L. and Caves, Carlton M.},
  journal       = {Physical Review Letters},
  volume        = {72},
  issue         = {22},
  pages         = {3439--3443},
  numpages      = {0},
  year          = {1994},
  month         = {May},
  publisher     = {American Physical Society},
  url           = {https://link.aps.org/doi/10.1103/PhysRevLett.72.3439}
}

@article{Passos2020,
  title = {Spin-orbit implementation of the Solovay-Kitaev decomposition of single-qubit channels},
  author = {Passos, M. H. M. and Junior, A. de Oliveira and de Oliveira, M. C. and Khoury, A. Z. and Huguenin, J. A. O.},
  journal = {Physical Review A},
  volume = {102},
  issue = {6},
  pages = {062601},
  numpages = {13},
  year = {2020},
  month = {Dec},
  publisher = {American Physical Society},
  doi = {10.1103/PhysRevA.102.062601},
  url = {https://link.aps.org/doi/10.1103/PhysRevA.102.062601}
}

@article{MacLellan2024,
  author        = {MacLellan, Benjamin and Roztocki, Piotr and Czischek, Stefanie and Melko, Roger G.},
  title         = {End-to-end variational quantum sensing},
  journal       = {npj Quantum Information},
  year          = {2024},
  month         = {Nov},
  day           = {19},
  volume        = {10},
  number        = {1},
  pages         = {118},
  issn          = {2056-6387},
  doi           = {10.1038/s41534-024-00914-w},
  url           = {https://doi.org/10.1038/s41534-024-00914-w}
}

@article{Altepeter2005,
  author        = {J. B. Altepeter and E. R. Jeffrey and P. G. Kwiat},
  journal       = {Optics Express},
  number        = {22},
  pages         = {8951--8959},
  publisher     = {Optica Publishing Group},
  title         = {Phase-compensated ultra-bright source of entangled photons},
  volume        = {13},
  month         = {Oct},
  year          = {2005},
  url           = {https://opg.optica.org/oe/abstract.cfm?URI=oe-13-22-8951}
}

@article{Akselrod2007,
  author        = {G. M. Akselrod and J. B. Altepeter and E. R. Jeffrey and P. G. Kwiat},
  journal       = {Optics Express},
  number        = {8},
  pages         = {5260--5261},
  publisher     = {Optica Publishing Group},
  title         = {Phase-compensated ultra-bright source of entangled photons: erratum},
  volume        = {15},
  month         = {Apr},
  year          = {2007},
  url           = {https://opg.optica.org/oe/abstract.cfm?URI=oe-15-8-5260}
}

@article{Rangarajan2009,
  author        = {Radhika Rangarajan and Michael Goggin and Paul Kwiat},
  journal       = {Optics Express},
  number        = {21},
  pages         = {18920--18933},
  publisher     = {Optica Publishing Group},
  title         = {Optimizing type-I polarization-entangled photons},
  volume        = {17},
  month         = {Oct},
  year          = {2009},
  url           = {https://opg.optica.org/oe/abstract.cfm?URI=oe-17-21-18920}
}

@article{Kaubruegger2023,
  title = {Optimal and Variational Multiparameter Quantum Metrology and Vector-Field Sensing},
  author = {Kaubruegger, Raphael and Shankar, Athreya and Vasilyev, Denis V. and Zoller, Peter},
  journal = {PRX Quantum},
  volume = {4},
  issue = {2},
  pages = {020333},
  numpages = {21},
  year = {2023},
  month = {Jun},
  publisher = {American Physical Society},
  doi = {10.1103/PRXQuantum.4.020333},
}

@article{Valeri2023,
  title = {Experimental multiparameter quantum metrology in adaptive regime},
  author = {Valeri, Mauro and others},
  journal = {Physical Review Research},
  volume = {5},
  issue = {1},
  pages = {013138},
  numpages = {11},
  year = {2023},
  month = {Feb},
  publisher = {American Physical Society},
  doi = {10.1103/PhysRevResearch.5.013138},
}

@article{Ferreira2024,
  author={Ferreira, Vinicius S. and Kim, Gihwan and Butler, Andreas and Pichler, Hannes and Painter, Oskar},
  title={Deterministic generation of multidimensional photonic cluster states with a single quantum emitter},
  journal={Nature Physics},
  year={2024},
  month={May},
  day={01},
  volume={20},
  number={5},
  pages={865-870},
  issn={1745-2481},
  doi={10.1038/s41567-024-02408-0},
  url={https://doi.org/10.1038/s41567-024-02408-0}
}

@article{Huet2025,
  author={Huet, H. and others},
  title={Deterministic and reconfigurable graph state generation with a single solid-state quantum emitter},
  journal={Nature Communications},
  year={2025},
  month={May},
  day={09},
  volume={16},
  number={1},
  pages={4337},
  issn={2041-1723},
  doi={10.1038/s41467-025-59693-3},
  url={https://doi.org/10.1038/s41467-025-59693-3}
}

@article{Qin2025,
  title = {Scaling of computational order parameters in Rydberg-atom graph states},
  author = {Qin, Zhangjie and Scarola, V. W.},
  journal = {Physical Review A},
  volume = {111},
  issue = {4},
  pages = {042617},
  numpages = {12},
  year = {2025},
  month = {Apr},
  publisher = {American Physical Society},
  doi = {10.1103/PhysRevA.111.042617},
  url = {https://link.aps.org/doi/10.1103/PhysRevA.111.042617}
}

@article{OSullivan2025,
  author={O'Sullivan, James and others},
  title={Deterministic generation of two-dimensional multi-photon cluster states},
  journal={Nature Communications},
  year={2025},
  month={Jul},
  day={01},
  volume={16},
  number={1},
  pages={5505},
  issn={2041-1723},
  doi={10.1038/s41467-025-60472-3},
  url={https://doi.org/10.1038/s41467-025-60472-3}
}

@article{Young2011,
  title = {Quantum-dot-induced phase shift in a pillar microcavity},
  author = {Young, A. B. and others},
  journal = {Physical Review A},
  volume = {84},
  issue = {1},
  pages = {011803},
  numpages = {4},
  year = {2011},
  month = {Jul},
  publisher = {American Physical Society},
  doi = {10.1103/PhysRevA.84.011803},
  url = {https://link.aps.org/doi/10.1103/PhysRevA.84.011803}
}

@article{Ringbauer2025,
  author={Ringbauer, Martin and others},
  title={Verifiable measurement-based quantum random sampling with trapped ions},
  journal={Nature Communications},
  year={2025},
  month={Jan},
  day={02},
  volume={16},
  number={1},
  pages={106},
  issn={2041-1723},
  doi={10.1038/s41467-024-55342-3},
  url={https://doi.org/10.1038/s41467-024-55342-3}
}

@article{Browne2005,
  title = {Resource-Efficient Linear Optical Quantum Computation},
  author = {Browne, Daniel E. and Rudolph, Terry},
  journal = {Physical Review Letters},
  volume = {95},
  issue = {1},
  pages = {010501},
  numpages = {4},
  year = {2005},
  month = {Jun},
  publisher = {American Physical Society},
  doi = {10.1103/PhysRevLett.95.010501},
  url = {https://link.aps.org/doi/10.1103/PhysRevLett.95.010501}
}

@article{Kwiat1999,
  title={Ultrabright source of polarization-entangled photons},
  volume={60},
  ISSN={1094-1622},
  url={http://dx.doi.org/10.1103/PhysRevA.60.R773},
  DOI={10.1103/physreva.60.r773},
  number={2},
  journal={Physical Review A},
  publisher={American Physical Society (APS)},
  author={Kwiat, Paul G. and Waks, Edo and White, Andrew G. and Appelbaum, Ian and Eberhard, Philippe H.},
  year={1999},
  month=aug, 
  pages={R773–R776}
}

@article{Moses2023,
  title = {A Race-Track Trapped-Ion Quantum Processor},
  author = {Moses, S. A. and others},
  journal = {Physical Review X},
  volume = {13},
  issue = {4},
  pages = {041052},
  numpages = {25},
  year = {2023},
  month = {Dec},
  publisher = {American Physical Society},
  doi = {10.1103/PhysRevX.13.041052},
  url = {https://link.aps.org/doi/10.1103/PhysRevX.13.041052}
}

@article{Pogorelov2021,
  title = {Compact Ion-Trap Quantum Computing Demonstrator},
  author = {Pogorelov, I. and others},
  journal = {PRX Quantum},
  volume = {2},
  issue = {2},
  pages = {020343},
  numpages = {23},
  year = {2021},
  month = {Jun},
  publisher = {American Physical Society},
  doi = {10.1103/PRXQuantum.2.020343},
  url = {https://link.aps.org/doi/10.1103/PRXQuantum.2.020343}
}

@article{Le2023,
       author = {{Le}, Trung Kien and {Nguyen}, Hung Q. and {Ho}, Le Bin},
        title = "{Variational quantum metrology for multiparameter estimation under dephasing noise}",
      journal = {Scientific Reports},
         year = 2023,
        month = oct,
       volume = {13},
          eid = {17775},
        pages = {17775},
          doi = {10.1038/s41598-023-44786-0},
}

@article{Toth2014,
  doi = {10.1088/1751-8113/47/42/424006},
  url = {https://doi.org/10.1088/1751-8113/47/42/424006},
  year = {2014},
  month = {oct},
  publisher = {IOP Publishing},
  volume = {47},
  number = {42},
  pages = {424006},
  author = {Tóth, Géza and Apellaniz, Iagoba},
  title = {Quantum metrology from a quantum information science perspective},
  journal = {Journal of Physics A: Mathematical and Theoretical},
}

@article{Kang2025,
  author={Kang, Haiyue and Kam, John F. and Mooney, Gary J. and Hollenberg, Lloyd C. L.},
  title={Entanglement teleportation along a regenerating hamster-wheel graph state},
  journal={Scientific Reports},
  year={2025},
  month={Dec},
  day={03},
  issn={2045-2322},
  doi={10.1038/s41598-025-30301-0},
  url={https://doi.org/10.1038/s41598-025-30301-0}
}

@article{Proctor2018,
  title = {Multiparameter Estimation in Networked Quantum Sensors},
  author = {Proctor, Timothy J. and Knott, Paul A. and Dunningham, Jacob A.},
  journal = {Phys. Rev. Lett.},
  volume = {120},
  issue = {8},
  pages = {080501},
  numpages = {6},
  year = {2018},
  month = {Feb},
  publisher = {American Physical Society},
  doi = {10.1103/PhysRevLett.120.080501},
  url = {https://link.aps.org/doi/10.1103/PhysRevLett.120.080501}
}

@article{Ge2018,
  title = {Distributed Quantum Metrology with Linear Networks and Separable Inputs},
  author = {Ge, Wenchao and Jacobs, Kurt and Eldredge, Zachary and Gorshkov, Alexey V. and Foss-Feig, Michael},
  journal = {Phys. Rev. Lett.},
  volume = {121},
  issue = {4},
  pages = {043604},
  numpages = {6},
  year = {2018},
  month = {Jul},
  publisher = {American Physical Society},
  doi = {10.1103/PhysRevLett.121.043604},
  url = {https://link.aps.org/doi/10.1103/PhysRevLett.121.043604}
}

@article{Zhang2021,
    doi = {10.1088/2058-9565/abd4c3},
    url = {https://doi.org/10.1088/2058-9565/abd4c3},
    year = {2021},
    month = {jul},
    publisher = {IOP Publishing},
    volume = {6},
    number = {4},
    pages = {043001},
    author = {Zhang, Zheshen and Zhuang, Quntao},
    title = {Distributed quantum sensing},
    journal = {Quantum Science and Technology},
}

@article{Kim2024,
    author={Kim, Dong-Hyun and others},
    title={Distributed quantum sensing of multiple phases with fewer photons},
    journal={Nature Communications},
    year={2024},
    month={Jan},
    day={11},
    volume={15},
    number={1},
    pages={266},
    issn={2041-1723},
    doi={10.1038/s41467-023-44204-z},
    url={https://doi.org/10.1038/s41467-023-44204-z}
}

@misc{Hein2006,
  author        = {Hein, M. and others},
  year          = {2006},
  eprint        = {quant-ph/0602096},
  archiveprefix = {arXiv},
  primaryclass  = {quant-ph},
  url           = {https://arxiv.org/abs/quant-ph/0602096}
}

@misc{Zhang2024,
      author={Zhang, Jiajian and others},
      year={2024},
      eprint={2412.18398},
      archivePrefix={arXiv},
      primaryClass={quant-ph},
      url={https://arxiv.org/abs/2412.18398},
}

@misc{Rimock2024,
  author        = {Noam Rimock and Khen Cohen and Yaron Oz},
  year          = {2024},
  eprint        = {2406.15666},
  archiveprefix = {arXiv},
  primaryclass  = {quant-ph},
  url           = {https://arxiv.org/abs/2406.15666}
}

@misc{Crooks2019,
  author        = {Gavin E. Crooks},
  year          = {2019},
  eprint        = {1905.13311},
  archiveprefix = {arXiv},
  primaryclass  = {quant-ph},
  url           = {https://arxiv.org/abs/1905.13311}
}

@inproceedings{Petz2011,
  title         = {Introudction to Quantum Fisher Information},
  url           = {http://dx.doi.org/10.1142/9789814338745_0015},
  booktitle     = {Quantum Probability and Related Topics},
  publisher     = {World Scientific},
  author        = {Petz, D. and Ghinea, C.},
  year          = {2011},
  month         = jan
}

@book{Holevo2011,
  author        = {Holevo, Alexander S},
  title         = {Probabilistic and Statistical Aspects of Quantum Theory},
  publisher     = {Springer},
  address       = {Dordrecht},
  series        = {Publications of the Scuola Normale Superiore. Monographs},
  year          = {2011},
  url           = {https://cds.cern.ch/record/1414149}
}

@book{Efron1993,
  author        = {Efron, Bradley and Tibshirani, Robert J},
  title         = {An introduction to the bootstrap},
  publisher     = {Chapman and Hall},
  address       = {London},
  series        = {Chapman \& Hall/CRC monographs on statistics and applied probability},
  year          = {1993},
  url           = {https://cds.cern.ch/record/526679}
}

@book{MikeIke2011,
  author = {Nielsen, Michael A. and Chuang, Isaac L.},
  title = {Quantum Computation and Quantum Information: 10th Anniversary Edition},
  year = {2011},
  isbn = {1107002176},
  publisher = {Cambridge University Press},
  address = {USA},
  edition = {10th},
}

@misc{Alexander2025,
      author={B. J. Alexander and S. K. Ozdemir and M. S. Tame},
      year={2026},
      eprint={2602.13026},
      archivePrefix={arXiv},
      primaryClass={quant-ph},
      url={https://arxiv.org/abs/2602.13026}, 
}

\appendix
\onecolumngrid
\newpage

\counterwithin{equation}{section}
\counterwithin{table}{section}
\counterwithin{figure}{section}

\section{Single-qubit rotation gates with polarization optics\label{sec:single_q_rotation_gates_with_pol_optics}}
A general $\text{SU}(2)$ matrix can be written as a sequence of three rotations
characterized by the Euler angles \((\varphi, \xi, \zeta)\),
\begin{align}
    \label{eq:general_su_2}
    U(\varphi, \xi, \zeta) = \tilde{R}_y(\varphi) \tilde{R}_z(-\xi) \tilde{R}_y(\zeta),
\end{align}
\noindent
where $\tilde{R}_{\hat{\mathbf{n}}}(\psi) = e^{-i\psi \hat{\mathbf{n}} \cdot
\mathbf{\sigma}}$. Here $\mathbf{\hat{n}}$ is a unit vector and $\mathbf{\sigma}
= (\sigma_x, \sigma_y, \sigma_z)$ are the Pauli matrices. Note that the
$\tilde{R}_{\hat{\mathbf{n}}}(\psi/2) = R_{\hat{\mathbf{n}}}(\psi)$, where the
latter is the standard definition~\cite{MikeIke2011}. For polarization-based
operations, a general $\text{SU}(2)$ operator can be decomposed as a product of
waveplate operations in the following way:
\begin{align}
    \label{eq:qhq_su_2}
    U(\eta_1, \tau, \eta_2) = \text{QWP}(\eta_1)\text{HWP}(\tau)\text{QWP}(\eta_2)
\end{align}
\noindent
where 
\begin{align}
    \label{eq:qwp_hwp_q0_h0}
    \text{QWP}(\eta) &= \tilde{R}_y(\eta) Q_0 \tilde{R}_y(-\eta), \\
    \text{HWP}(\tau) &= \tilde{R}_y(\tau) H_0 \tilde{R}_y(-\tau), \nonumber \\
    Q_0 &= \mqty[1 & 0 \\ 0 & i], \nonumber \\
    H_0 &= \mqty[1 & 0 \\ 0 & -1]. \nonumber
\end{align}

A simple relation can be established between the waveplates and the Euler
angles~\cite{Passos2020}:
\begin{align}
    \label{eq:euler_wavaplates_angles}
    \eta_1 &= \varphi - \pi/4, \\
    \eta_2 &= -\zeta - \pi/4, \nonumber \\
    \tau &= (\varphi + \xi - \zeta)/2 - \pi/4. \nonumber
\end{align}

\subsection{Waveplate angles for x-axis rotation operator}
We obtain $R_{x}(\theta)$ by setting $\varphi=-\pi/4$, $\zeta = \pi/4$ and $\xi=\theta/2$ in Eq.~\ref{eq:general_su_2} to give
\begin{align}
    U(-\pi/4, \theta/2, \pi/4) = \tilde{R}_y(-\pi/4)\tilde{R}_z(-\theta/2)\tilde{R}_y(\pi/4) = \tilde{R}_x(\theta/2) = R_{x}(\theta).
\end{align}
\noindent
The relation between the waveplate angles and $\theta$ are given by

\begin{align}
    \label{eq:waveplates_rx}
    \eta_1 &= -\pi/2, \\
    \eta_2 &= -\pi/2, \nonumber \\
    \tau &= \theta/4 - \pi/2. \nonumber
\end{align}

\subsection{Waveplate angles for y-axis rotation operator}
Similarly, $R_y(\theta)$ can be obtained by setting $\varphi=\xi=0$ and $\zeta = \theta/2$ in Eq.~\ref{eq:general_su_2} to give
\begin{align}
    U(0, 0, \theta/2) = \tilde{R}_y(0)\tilde{R}_z(0)\tilde{R}_y(\theta/2) = \tilde{R}_y(\theta/2) = R_{y}(\theta),
\end{align}
\noindent
where the waveplate angles are related to $\theta$ via

\begin{align}
    \label{eq:waveplates_ry}
    \eta_1 &= - \pi/4, \\
    \eta_2 &= -\theta/2 - \pi/4, \nonumber \\
    \tau &= -\theta/4 - \pi/4. \nonumber
\end{align}

\subsection{Waveplate angles for z-axis rotation operator}
Setting $\varphi=\zeta=0$ and $\xi=-\theta/2$, Eq.~\ref{eq:general_su_2} becomes:
\begin{align}
    U(0, -\theta/2, 0) = \tilde{R}_y(0)\tilde{R}_z(\theta/2)\tilde{R}_y(0) = \tilde{R}_z(\theta/2) = R_z(\theta).
\end{align}
\noindent
The waveplate angles are related to $\theta$ via
\begin{align}
    \label{eq:waveaples_rz}
    \eta_1 &= -\pi/4, \\
    \eta_2 &= -\pi/4, \nonumber \\
    \tau &= -\theta/4 - \pi/4. \nonumber
\end{align}

\subsection{Waveplate angles for identity operator}
Setting  $\varphi=\zeta=\xi=0$, Eq.~\ref{eq:general_su_2} becomes:
\begin{align}
    U(0, 0, 0) = \tilde{R}_y(0)\tilde{R}_z(0)\tilde{R}_y(0) = I,
\end{align}
\noindent
where waveplate angles are given by

\begin{align}
    \label{eq:waveaples_id}
    \eta_1 &= -\pi/4, \\
    \eta_2 &= -\pi/4, \nonumber \\
    \tau &= -\pi/4. \nonumber
\end{align}
\noindent
The angles in Passos et al.~\cite{Passos2020} are for the fast axis, measured from the
horizontal and positive in the anticlockwise direction, when the beam passes
the label‐side first.  To convert to an angle $\theta$ of the fast axis
measured from the vertical, positive in the clockwise direction for a beam
entering the label‐side first, we use $ \tau_{\text{lab}} = \pi/2 -
\tau_{\text{Passos}}$.

\section{Fringe visibility \label{sec:fringe_visibility}}
\begin{figure}[H]
    \centering
    \includegraphics[width=0.75\linewidth]{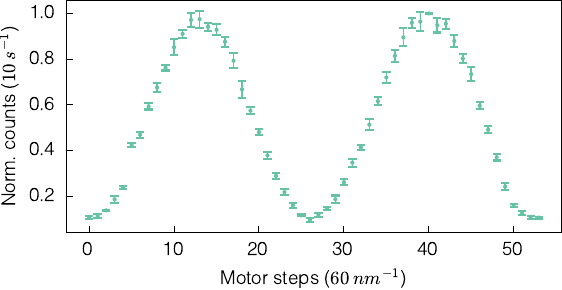}
    \caption{
	Normalized photon counts as a function of the phase difference between
	the two paths, varied by translating the mirror on a stage in steps of
	$60$ nm (forward). Each data point is the average of 10 measurements.
	The error bars represent the standard deviation of the mean. The
	measurement setting was
	$\ket*{A_1}\bra*{A_1}\otimes\ket*{H_2}\bra*{H_2}$. The jump around the
	6th step was due to jogging twice (120 nm step).
    }
    \label{fig:fringe-visibility-norm-ah}
\end{figure}

The normalized photon counts as a function of motor steps in
Fig.~\ref{fig:fringe-visibility-norm-ah} are fitted to the function $f(\varphi)
= a\cos\left(b\varphi + c\right) + d$, where the parameters of the fit are $a,
b, c$ and $d$. We find the following values for the fit parameters: $a =
-0.445$, $b = 0.235$, $c = 0.0662$ and $d = 0.531$ to reproduce
Fig.~\ref{fig:bell_state_fv_ah_fit} in the main text.

\section{Polarization measurement bases \label{sec:polarization_measurement_bases}}
We follow the conventions in Ref.~\cite{James2001} but deviate slightly as they
consider a linear polarizer transmits vertically polarized light, while we
consider a polarizing-beam-splitter that transmits horizontally polarized
light. Moreover, our waveplate angles are always clockwise with respect to the
vertical axis and fast axis of the waveplates. The 16 projective measurements
we use in our tomography experiments and their waveplate angles are shown in
Tab.~\ref{tab:2q_tomo_measurements}. Moreover, we use a Monte Carlo simulation
with Poisson noise as the main source of noise~\cite{James2001} where we
generate 100 density matrices, and derive from these the average and standard
deviation of the quantities of interest.

\begin{table}[H]
 	\centering
	\begin{tabular}{ccccccc}
		\toprule
		$\hat{P}_{1} \otimes \hat{P}_{2}$ & $h_1$ & $q_1$ & $h_2$ & $q_2$ \\
		\toprule
		 $\op*{V}\otimes\op*{V}$ & $45^{\circ}$   & $0^{\circ}$  & $45^{\circ}$    & $0^{\circ}$   \\
		 $\op*{V}\otimes\op*{H}$ & $45^{\circ}$   & $0^{\circ}$  & $0^{\circ}$     & $0^{\circ}$   \\
		 $\op*{H}\otimes\op*{H}$ & $0^{\circ}$    & $0^{\circ}$  & $0^{\circ}$     & $0^{\circ}$   \\
		 $\op*{H}\otimes\op*{V}$ & $0^{\circ}$    & $0^{\circ}$  & $45^{\circ}$    & $0^{\circ}$   \\
		 $\op*{L}\otimes\op*{V}$ & $-22.5^{\circ}$ & $0^{\circ}$  & $45^{\circ}$    & $0^{\circ}$   \\
		 $\op*{L}\otimes\op*{H}$ & $-22.5^{\circ}$ & $0^{\circ}$  & $0^{\circ}$     & $0^{\circ}$   \\
		 $\op*{D}\otimes\op*{H}$ & $-22.5^{\circ}$ & $45^{\circ}$ & $0^{\circ}$     & $0^{\circ}$   \\
		 $\op*{D}\otimes\op*{V}$ & $-22.5^{\circ}$ & $45^{\circ}$ & $45^{\circ}$    & $0^{\circ}$   \\
		 $\op*{D}\otimes\op*{L}$ & $-22.5^{\circ}$ & $45^{\circ}$ & $-22.5^{\circ}$  & $0^{\circ}$   \\
		 $\op*{D}\otimes\op*{D}$ & $-22.5^{\circ}$ & $45^{\circ}$ & $-22.5^{\circ}$  & $45^{\circ}$  \\
		 $\op*{L}\otimes\op*{D}$ & $-22.5^{\circ}$ & $0^{\circ}$  & $-22.5^{\circ}$  & $45^{\circ}$  \\
		 $\op*{V}\otimes\op*{D}$ & $45^{\circ}$   & $0^{\circ}$  & $45^{\circ}$    & $45^{\circ}$  \\
		 $\op*{H}\otimes\op*{D}$ & $0^{\circ}$    & $0^{\circ}$  & $-22.5^{\circ}$     & $45^{\circ}$  \\
		 $\op*{H}\otimes\op*{R}$ & $0^{\circ}$    & $0^{\circ}$  & $22.5^{\circ}$     & $0^{\circ}$  \\
		 $\op*{V}\otimes\op*{R}$ & $45^{\circ}$   & $0^{\circ}$  & $22.5^{\circ}$    & $0^{\circ}$  \\
		 $\op*{L}\otimes\op*{R}$ & $-22.5^{\circ}$ & $0^{\circ}$  & $22.5^{\circ}$  & $0^{\circ}$  \\
		\toprule
	\end{tabular}
	\caption{
	    Measurement settings for the tomography analysis of the two-photon
	    polarization state. Coincidence counts are collected over a 10 second
	    collection window. Here, $\ket*{D}\ =(\ket*{H}+\ket*{V})/\sqrt{2}$,
	    $\ket*{A} = (\ket*{H}-\ket*{V})/\sqrt{2}$, $\ket*{L} =
	    (\ket*{H}+i\ket*{V})/\sqrt{2}$ and $\ket*{R} =
	    (\ket*{H}-i\ket*{V})/\sqrt{2}$. The angles $h$ and $q$ denote angles for
	    the HWPs and QWPs, respectively.
	}
        \label{tab:2q_tomo_measurements}
\end{table}

\section{Closed-form expression for the quantum Fisher information \label{sec:closed_form_qfi}}
For a pure state $\rho$ and a generating Hamiltonian operator $\hat{H}$, the
quantum Fisher information can be written in terms of the single-shot variance
of $\hat{H}$~\cite{Paris2009} as
\begin{align}
    F_{\mathcal{Q}}(\rho, \hat{H}) = 4 (\Delta H)^2,
\end{align}
\noindent
where $(\Delta H)^2 = \expval*{\hat{H}^2} - \expval*{\hat{H}}^2$. In our particular case, $\hat{H}$ is given by
\begin{align}
    \hat{H} = \frac{1}{2}(\hat{X}_1\otimes\mathds{1}_2 + \mathds{1}_1\otimes\hat{Z}_{2}).
\end{align}
\noindent
The expectation value of the above Hamiltonian operator with respect to the
weighted graph state $\ket*{\Gamma_{\phi_{12}}} = \mqty[1 & 1 & 1 &
e^{i\phi_{12}}]^{\mathsf{T}}/2$ can be written as
\begin{align}
    \expval*{\hat{H}} = \expval*{\frac{1}{2} (\hat{X}_1\otimes\mathds{1}_{2} + \mathds{1}_{1}\otimes\hat{Z}_2)} = \frac{1}{2} (\expval*{\hat{X}_1\otimes\mathds{1}_{2}} + \expval*{\mathds{1}_1\otimes\hat{Z}_2}).
     \label{eq:h_gen_expval}
\end{align}
\noindent
The first expectation value becomes:
\begin{align}
    \expval*{\hat{X}_1\otimes\mathds{1}_{2}} &=  \ev*{\hat{X}_{1}\otimes\mathds{1}_{2}}{\Gamma_{\phi_{12}}} \nonumber, \\
    &= \frac{1}{4} (\bra*{00} + \bra*{01} + \bra*{10} + e^{-i\phi_{12}}\bra*{11}) \cdot (\ket*{10} + \ket*{11} + \ket*{00} + e^{i\phi_{12}}\ket*{01}) \nonumber, \\
    &= \frac{1}{4} (\braket*{00}{00} + e^{i\phi_{12}}\braket*{01}{01} + \braket*{10}{10} + e^{-i\phi_{12}}\braket*{11}{11}) \nonumber, \\
    &= \frac{1}{4} (1 + e^{i\phi_{12}} + 1 + e^{-i\phi_{12}}) = \frac{1}{4} (2 + 2\cos\phi_{12})  = \frac{1}{2}(1 + \cos\phi_{12}).
\end{align}
\noindent
The second expectation value in Eq.~\ref{eq:h_gen_expval} becomes,
\begin{align}
    \expval*{\mathds{1}_{1}\otimes\hat{Z}_2} &= \ev*{\mathds{1}_{1}\otimes\hat{Z}_2}{\Gamma_{\phi_{12}}} \nonumber, \\
    &= \frac{1}{4} (\bra*{00} + \bra*{01} + \bra*{10} + e^{-i\phi_{12}}\bra*{11}) \cdot (\ket*{00} - \ket*{01} + \ket*{10} - e^{i\phi_{12}}\ket*{11}) \nonumber, \\
    &= \frac{1}{4} (\braket*{00}{00} - \braket*{01}{01} + \braket*{10}{10} - e^{i\phi_{12}}e^{-i\phi_{12}}\braket*{11}{11}) \nonumber, \\
    &= \frac{1}{4} (1 - 1 + 1 - 1) = 0.
\end{align}
\noindent
Putting everything together, Eq.~\ref{eq:h_gen_expval} becomes,
\begin{align}
    \ev*{\hat{H}} = \frac{1}{2}\left(\frac{1}{2}(1 + \cos{\phi_{12}}) + 0\right) = \frac{1}{4}(1 + \cos{\phi_{12}}).
\end{align}
\noindent
To calculate $\ev*{\hat{H}^2}$, we first write $\hat{H}^2$,
\begin{align}
    \hat{H}^2 &= \left[ \frac{1}{2}(\hat{X}_1\otimes\mathds{1}_{2} + \mathds{1}_{1}\otimes\hat{Z}_2) \right]^2 = \frac{1}{4} (\hat{X}_1^2\otimes\mathds{1}_{2} + \mathds{1}_{1}\otimes\hat{Z}_2^2 +  2\hat{X}_1\otimes\hat{Z}_2).
\end{align}
\noindent
Since $\hat{X}^2 = \hat{Z}^2 = \mathds{1}$,

\begin{align}
	\hat{H}^2 = \frac{1}{4} (2\mathds{1}_{1}\otimes\mathds{1}_{2} + 2\hat{X}_1\otimes\hat{Z}_2) = \frac{1}{2} (\mathds{1}_{1}\otimes\mathds{1}_2 + \hat{X}_1\otimes\hat{Z}_2).
\end{align}
\noindent
The expectation value is:
\begin{align}
    \ev*{\hat{H}^2} = \frac{1}{2} (\ev*{\mathds{1}_{1}\otimes\mathds{1}_{2}} + \ev*{\hat{X}_1\otimes\hat{Z}_2}) = \frac{1}{2} (1 + \ev*{\hat{X}_1\otimes\hat{Z}_2}).
    \label{eq:h_gen_expval_2}
\end{align}
\noindent
The second expectation value in the expression evaluates to 
\begin{align}
    \ev*{\hat{X}_1\otimes\hat{Z}_2} &= \ev*{\hat{X}_1\otimes\hat{Z}_2}{\Gamma_{\phi_{12}}} \nonumber, \\
    &= \frac{1}{4} (\bra*{00} + \bra*{01} + \bra*{10} + e^{-i\phi_{12}}\bra*{11}) \cdot (\ket*{10} - \ket*{11} + \ket*{00} - e^{i\phi_{12}}\ket*{01}) \nonumber, \\
    &= \frac{1}{4} (\braket*{00}{00} + -e^{i\phi_{12}}\braket*{01}{01} + \braket*{10}{10} - e^{-i\phi_{12}}\braket*{11}{11}) \nonumber, \\
    &= \frac{1}{4} (1 - e^{i\phi_{12}} + 1 - e^{-i\phi_{12}}) = \frac{1}{4} (2 - 2\cos\phi_{12}) = \frac{1}{2}(1 - \cos\phi_{12}).
\end{align}
\noindent
Collecting everything, Eq.~\ref{eq:h_gen_expval_2} becomes
\begin{align}
    \ev*{\hat{H}^{2}}  = \frac{1}{2}\left(1 + \frac{1}{2}(1 - \cos\phi_{12})\right) = \frac{1}{4}(3 - \cos{\phi_{12}}).
\end{align}
\noindent
The single-shot variance for $\hat{H}$ is then given by
\begin{align}
    (\Delta \hat{H})^2 &= \ev*{\hat{H}^2} - \ev*{\hat{H}}^2 \nonumber, \\
		       &= \frac{1}{4}(3 - \cos{\phi_{12}}) - \left[\frac{1}{4}(1  + \cos{\phi_{12}})\right]^2 \nonumber, \\
		    &= \frac{4(3 - \cos\phi_{12})}{16} - \frac{1 + 2\cos\phi_{12} + \cos^2\phi_{12}}{16} \nonumber, \\
		    &= \frac{12 - 4\cos\phi_{12} - 1 - 2\cos\phi_{12} - \cos^2\phi_{12}}{16} \nonumber, \\
		    &= \frac{11 - 6\cos\phi_{12} - \cos^2\phi_{12}}{16}.
\end{align}
\noindent
To get the expression for the QFI, we multiply the expression above by 4
\begin{align}
    F_{\mathcal{Q}}(\dyad*{\Gamma_{\phi_{12}}}, \hat{H}) = \frac{11 - 6\cos\phi_{12} - \cos^2{\phi_{12}}}{4},
\end{align}
\noindent
the QFI is independent of the sensing phase $\theta$ and depends only on the
graph weight $\phi_{12}$. When $\phi_{12}=\pi$, we have $F_{\mathcal{Q}}=4$,
which is the HL for $N=2$, on the other hand, when $\phi_{12}=0$ we have
$F_{\mathcal{Q}}=1$, which is below the SQL for $N=2$ ($F_{\mathcal{Q}}=2$).
Therefore, a range of enhancements in the precision $(\Delta \theta)^2$ is
expected depending on the weight $\phi_{12}$, according to the QCRB.

\section{Local Pauli measurements \label{sec:local_pauli_measurements}}
\begin{figure}[h]
    \centering
      \includegraphics[width=\linewidth]{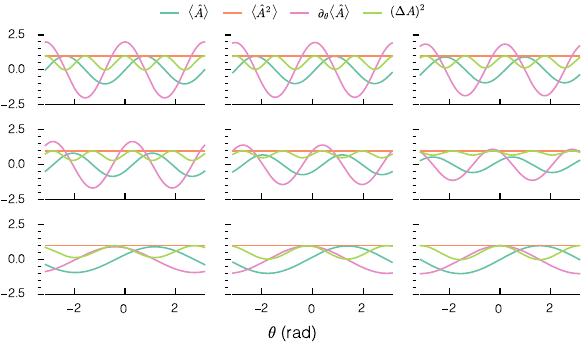}
      \caption{
	  Top left to bottom right: The expectation value of
	  $\expval*{\hat{A}}$ (blue line), its square $\expval*{\hat{A}^2}$
	  (orange line), derivative $\partial_{\theta}\expval*{\hat{A}}$ (green
	  line) and variance $(\Delta A)^2$ (red line)  as a function of the
	  encoded phase $\theta \in [0, 2\pi)$ for graph weights $\phi_{12} \in
	  \{\pi, \tfrac{7\pi}{8}, \tfrac{3\pi}{4}, \tfrac{5\pi}{8},
	  \tfrac{\pi}{2}, \tfrac{3\pi}{8}, \tfrac{\pi}{4}, \tfrac{\pi}{8},
      0\}$. Here, $\hat{A}=\hat{A}_{1}\otimes\hat{A}_2$ is the optimal local
      Pauli operator at the phase encoding $\theta = 0$. These operators can be
      found in Tab.~\ref{tab:discrete_local_pauli}.
      }
      \label{fig:local_pauli_behaviour}
\end{figure}

We perform an exhaustive numerical search over all $4^2=16$ possible local
Pauli product operators at each graph weight $\phi_{12}$ of interest to find
the operator $\hat{A}=\hat{A}_{1}\otimes\hat{A}_2$ that yields the smallest
value for the estimator variance $(\Delta \theta)^2$.
Tab.~\ref{tab:discrete_local_pauli} shows the optimal local Pauli operators for
various graph weights $\phi_{12}$, when $\theta=0$. Here, when multiple
operators have same value for $(\Delta \theta)^2$ we choose one with
the largest derivative magnitude. Near the operating point $\theta=0$, the
derivative and variance of the operators $\hat{A}$ are robust enough to changes
in $\theta$. Particularly, the derivative reaches a maxima around
$\theta=0$ and $\expval*{\hat{A}}$ has an inflection point around $\theta=0$.
This is justification for approximating the derivative here via finite
differences. Fig.~\ref{fig:local_pauli_behaviour} shows the expectation value
of $\expval*{\hat{A}}$, its square $\expval*{\hat{A}^2}$, derivative
$\partial_{\theta}\expval*{\hat{A}}$ and variance $(\Delta A)^2$ of the optimal
local Pauli operators as a function of the encoded phase $\theta \in [0, 2\pi)$
for different graph weights.

\begin{table}[h]
    \centering
    \begin{tabular}{lcccc}
        \toprule
	$\phi_{12}$ & $\hat{A}$ & $\expval*{\hat{A}}$ & $\abs*{\partial_{\theta}\expval*{\hat{A}}}$ & $(\Delta \theta)^2$ \\
        \midrule
        $\pi$ & $\hat{Z}_1 \otimes \hat{Y}_{2}$ & 0.00 & 2.00 & 0.25 \\
        $7\pi/8$ & $\hat{Z}_1 \otimes \hat{Y}_{2}$ & -0.19 & 1.92 & 0.26 \\
        $3\pi/4$ & $\hat{Z}_1 \otimes \hat{Y}_{2}$ & -0.35 & 1.71 & 0.30 \\
        $5\pi/8$ & $\hat{Z}_1 \otimes \hat{Y}_2$ & -0.46 & 1.38 & 0.41 \\
        $\pi/2$ & $\hat{Z}_1 \otimes \hat{Y}_2$ & -0.50 & 1.0 & 0.75 \\
        $3\pi/8$ & $\hat{Y}_1 \otimes \hat{Y}_2$ & 0.31 & 0.92 & 1.06 \\
	$\pi/4$ & $\mathds{1}_1 \otimes \hat{Y}_2$ & 0.35 & 0.85 & 1.20 \\
	$\pi/8$ & $\mathds{1}_1 \otimes \hat{Y}_2$ & 0.19 & 0.96 & 1.04 \\
	$0$ & $\mathds{1}_1 \otimes \hat{Y}_2$ & 0.00 & 1.00 & 1.00 \\
        \bottomrule
    \end{tabular}
    \caption{Expectation values and $(\Delta \theta)^2$ obtained with optimal local Pauli measurements for varied weightings $\phi_{12}$.}
    \label{tab:discrete_local_pauli}
\end{table}

\newpage

\section{Bootstrapping \label{sec:bootstrapping}}
For a particular observable $\hat{A}$, we measure four combinations of
polarization projectors, labeled $k\in\{++, +-, -+, --\}$, with corresponding
weights $w_k \in \{ \pm 1 \}$ fixed by the chosen basis. For example, the
observable $\hat{A} = \hat{Z}_{1}\otimes\hat{Z}_2$ has the weights
$w_{++}=w_{--}=+1$, $w_{+-}=w_{-+}=-1$, while
$\hat{A} = \mathds{1}_{1}\otimes\hat{Y}_{2}$ has the weights
$w_{++}=w_{-+}=+1$ and $w_{+-}=w_{--}=-1$. Let $N_k$ be the observed
coincidences in one acquisition and $N=\sum_k N_k$ the total coincidences for
that setting. The estimator of the expectation value $\expval*{\hat{A}}$ is
\begin{align}
  \widehat{E} = \frac{\sum_k w_k N_k}{\sum_k N_k} = \frac{1}{N}\sum_k w_k N_k .
\end{align}
\noindent
Our data are naturally grouped into acquisition bins (runs) indexed by
$\ell=1,\ldots,L$, where $L=6$. Each bin contains the four photon counts
$N_{k,\ell}$ collected over a 10-second interval. One bootstrap replicate
proceeds as follows (and is performed independently for every setting used
below):
\begin{enumerate}
  \item Draw, with replacement, $L$ bin indices $\ell_1,\ldots,\ell_L\in\{1,\ldots,L\}$.
  \item Form resampled totals $N_k^{\star}=\sum_{j=1}^{L}N_{k,\ell_j}$ and $\nu^{\star}=\sum_k N_k^{\star}$.
  \item Set $\widehat{E}^{\star}=\big[\sum_k w_k N_k^{\star}\big]/\nu^{\star}$.
\end{enumerate}
Repeating for $b=1,\ldots,\mu$ gives the set
$\{\widehat{E}^{\star}_b\}_{b=1}^{\mu}$ and $\{1 -
(\widehat{E}^{\star}_b)^2\}_{b=1}^{\mu}$. We report point estimates as means of
the bootstrap samples and $95\%$ confidence intervals via the empirical $2.5$th
and $97.5$th percentiles.

\subsection{Poisson noise}
Note that we are interested in the single-shot variance $(\Delta A)^{2}$, where
the noise for a single pair of entangled photons in the state
$\ket*{\Gamma_{\phi_{12}}}$ originates from quantum fluctuations only. However,
in our experiment we are measuring coincidence counts corresponding to many
pairs of entangled photons over a fixed period of 10 seconds, where the number
of shots (photon pairs) fluctuates due to Poisson noise from the SPDC process.
This is equivalent to having a sample size $\nu$ that varies. Let $\nu^*$ be
the varying sample size due to Poisson noise, i.e., $\nu^* = \nu + \delta\nu$,
where $\nu$ is some fixed number of shots and $\delta_\nu \sim \mathcal{N}(0,
\sigma_\nu^2)$ is a Gaussian noise term with mean of zero and variance
$\sigma_\nu^2$. The expectation value of the variance of the estimator
$\hat{E}$, i.e., $(\Delta \hat{E})^2$, is then given by:
\begin{align}
    E[(\Delta \hat{E})^2] = E\left[\frac{(\Delta A)^2}{\nu^*}\right].
\end{align}
\noindent
Taking $(\Delta A)^2$ and $\nu^*$ to be independent, we have:
\begin{align}
    E\left[\frac{(\Delta A)^2}{\nu^*}\right] = E\left[\frac{1}{\nu^*}\right] E[(\Delta A)^2].
\end{align}
\noindent
The first term on the right hand side is:
\begin{align}
    E\left[\frac{1}{\nu^*}\right] = E\left[\frac{1}{\nu} \frac{1}{1 + \delta\nu/\nu}\right] \simeq E\left[\frac{1}{\nu}\right] = \frac{1}{\nu},
\end{align}
\noindent
where the final approximation is due to $\delta_\nu \approx \sqrt{\nu}$ for
Poisson noise and $\delta\nu/\nu = 1/\sqrt{\nu} \ll 1$ for large $\nu$.
Therefore we have:
\begin{align}
    E[(\Delta A)^2] = \nu E[(\Delta \hat{E})^2].
\end{align}
\noindent
The above relation shows that for large $\nu$, the measured variance from the
bootstrap procedure, $E[(\Delta \hat{E})^2]$, coming from $\mu$ samples, i.e.,
\begin{align}
    E[(\Delta \hat{E})^2] = \frac{1}{\mu} \sum_{b=1}^{\mu} \left(1 - (\hat{E}_b^*)^2\right),
\end{align}
\noindent
can be used to obtain the single-shot variance $E[(\Delta A)^2]$, using the
average count $\nu = \frac{1}{\mu} \sum_{b=1}^{\mu} \nu_b^*$.
The same approach can be followed for other expectation values, such as
$E[\hat{E}]$.

\subsection{Bootstrap for the 2-point finite-difference derivative}
The 2-point finite-difference derivative at $\theta^\ast$ with a fixed shift
$h>0$ ($h=5^{\circ}$ in our analysis) is estimated from
measurements at $\theta^\ast \pm h$ as
\begin{align}
    \eval{\partial_{\theta}\widehat{\expval*{\hat A}}}_{\theta=\theta^{\ast}}
    \approx \frac{\widehat{E}(\theta^{\ast} + h)-\widehat{E}(\theta^{\ast} - h)}{2h} .
  \label{eq:central-2pt}
\end{align}
At each setting $s \in \{\theta^{\ast} - h,\theta^{\ast} + h\}$, we perform the
bin bootstrap above to obtain $\{\widehat{E}^{\star}_b(s)\}_{b=1}^{\mu}$. For
each replicate $b$, define
\begin{equation}
  \partial_{\theta}\widehat{\expval*{\hat A}}^{\star}_b
  = \frac{\widehat{E}^{\star}_b(\theta^{\ast}{+}h)-\widehat{E}^{\star}_b(\theta^{\ast}{-}h)}{2h} .
  \label{eq:deriv-bootstrap}
\end{equation}
The point estimate and the $95\%$ CI are the mean and the percentile interval of
$\{\partial_{\theta} \widehat{\expval*{\hat A}}^{\star}_b\}_{b=1}^{\mu}$.

\subsection{Bootstrap for the estimator variance}
Consider the ratio for the estimator variance 
\begin{align}
    \widehat{\mathcal{R}}(\theta)
    = \frac{1 - \widehat{E}(\theta)^2}{\abs*{\partial_{\theta}\widehat{\expval*{\hat A}}}^2} .
  \label{eq:ratio}
\end{align}
To obtain the uncertainty for $\widehat{\mathcal{R}}(\theta)$, we use the same
bin bootstrap method independently at $\theta^\ast$, $\theta^\ast - h$ and
$\theta^\ast + h$. For each bootstrap replicate $b$,
\begin{align}
    \widehat{\mathcal{R}}^{\,\star}_b(\theta)
    =  \frac{1 - \big(\widehat{E}^{\star}_b(\theta^\ast)\big)^2}{
        \abs*{\partial_{\theta}\widehat{\expval*{\hat{A}}}^{\star}_b}^2} ,
\end{align}
with a small, fixed threshold applied to the denominator to avoid numerical
artefacts when the derivative is very small:
\begin{align}
    \abs*{\partial_{\theta}\widehat{\expval*{\hat{A}}}^{\star}_b}^2 \ \leftarrow \
    \max\!\big(\,\abs*{\partial_{\theta}\widehat{\expval*{\hat A}}^{\star}_b}^2,\ \varepsilon\,\big) ,
\end{align}
with $\varepsilon$ fixed \emph{a priori} (we used $\varepsilon\sim10^{-12}$). We
report the mean of $\{\widehat{\mathcal{R}}^{\,\star}_b(\theta)\}_{b=1}^{\mu}$ as the
point estimate and the $95\%$ percentile interval as its confidence interval.

\section{Local general-axis measurements \label{sec:general_axis_measurements}}

The central task is to find the optimal two-qubit observable, $\hat{A}$, that
minimizes the estimator variance, where $\hat{A}$ is a product of two local
single-qubit observables. Each local observable ($\hat{A}_{1}$ and
$\hat{A}_{2}$) is defined by its corresponding axis on the Bloch sphere using
two angles: the polar angle $\beta$ and the azimuthal angle $\alpha$. The
optimization routine finds the four optimal parameters: $(\beta_{1},
\alpha_{1}, \beta_{2}, \alpha_{2})$. The search is performed using a global
optimization technique called Differential Evolution, which explores the
four-dimensional parameter space. This is often followed by a local refinement
step using Powell's method to fine-tune the minimum. Both algorithms are
available in the Python library SciPy. We also add a small soft penalty term proportional to
$1/\abs*{\partial_{\theta}\expval*{\hat{A}}}$ to the objective function.
This ensures the optimizer favors solutions that not only yield low variance
but also maintain a high, measurable signal for the derivative. Moreover, after
optimization, the routine samples a small neighborhood around the best solution
and re-ranks these near-optimal points. This is done by first selecting all points
whose variance is within a small tolerance of the minimum, and from that
subset, we choose the point with the largest derivative
$\abs*{\partial_{\theta}\expval*{\hat{A}}}$. To physically realize the optimal local general-axis measurements, we need to realize the projectors for these
corresponding operators using a QWP, HWP, and PBS that only transmits
horizontally polarized photons. To find the waveplate angles corresponding
to each projector, we numerically minimize the objective function
$F_{\pm}(h,q)$
\begin{align}
    F_{\pm}(h^{(\pm)}, q^{(\pm)}) = 1 - \mel*{H}{\text{HWP}(h^{(\pm)}) \text{QWP}(q^{(\pm)})}{\beta, \alpha, \pm}.
\end{align}
\noindent
Here, $F_{\pm}(h, q)$ will reach a minimum when $\bra*{H}\text{HWP}(h^{(\pm)})
\text{QWP}(q^{(\pm)})  = \bra*{\pm, \alpha, \beta}$, \emph{i.e.}, acting with
$\text{HWP}(h^{(\pm)}) \text{QWP}(q^{(\pm)})$ on $\ket*{\beta, \alpha, \pm}$
maps it to $\ket*{H}$ (transmission port of the polarizing beam splitter). The
results of the optimization routines are shown in
Tab.~\ref{tab:general_axis_params}, which  shows the waveplate angles for each
projector, i.e., $++, +-, -+, --$ and the corresponding polar and
azimuthal angles. 

\begin{table}[h!]
  \centering
  \setlength{\tabcolsep}{3pt} 
  \resizebox{\textwidth}{!}{  
  \begin{tabular}{@{}c cccccc cccccc ccc@{}}
    \toprule
    $\phi_{12}$
    & $\beta_{1}$ & $\alpha_{1}$ & $h_1^{(+)}$ & $q_1^{(+)}$ & $h_1^{(-)}$ & $q_1^{(-)}$ 
    & $\beta_{2}$ & $\alpha_{2}$ & $h_2^{(+)}$ & $q_2^{(+)}$ & $h_2^{(-)}$ & $q_2^{(-)}$ 
    & $\ev*{\hat{A}}$ & $|\partial_{\theta}\ev*{\hat{A}}|$ & $(\Delta \theta)^2$ \\
    \midrule
    $\pi$    
    & $44.64$ & $-89.76$ & $-78.89$ & $-0.12$ & $-56.23$ & $89.88$  
    & $90.00$ & $44.60$  & $-11.37$ & $45.00$ & $-56.37$ & $45.00$ 
    & $0.01$ & $2.00$ & $0.25$ \\
    
    $7\pi/8$ 
    & $52.96$ & $-95.72$ & $-74.97$ & $3.76$  & $-56.27$ & $-86.24$ 
    & $85.73$ & $-140.95$& $-78.61$ & $-47.75$& $75.86$  & $42.25$ 
    & $0.24$ & $1.92$ & $0.26$ \\
    
    $3\pi/4$ 
    & $60.44$ & $-102.31$& $-70.30$ & $10.30$ & $-54.40$ & $-79.70$ 
    & $81.59$ & $-147.18$& $-78.10$ & $-49.99$& $73.11$  & $40.01$ 
    & $0.46$ & $1.69$ & $0.28$ \\
    
    $5\pi/8$ 
    & $64.54$ & $-110.61$& $84.70$  & $-71.76$& $68.54$  & $18.24$  
    & $79.36$ & $-159.05$& $24.80$  & $39.31$ & $69.80$  & $39.31$ 
    & $0.53$ & $1.50$ & $0.32$ \\
    
    $\pi/2$  
    & $68.27$ & $-120.53$& $89.68$  & $-64.06$& $71.36$  & $25.99$  
    & $79.26$ & $-174.10$& $21.25$  & $39.60$ & $63.35$  & $-50.40$
    & $0.55$ & $1.34$ & $0.39$ \\
    
    $3\pi/8$ 
    & $73.05$ & $-131.87$& $-84.99$ & $-57.27$& $72.72$  & $32.72$  
    & $81.22$ & $168.72$ & $17.47$  & $40.52$ & $68.05$  & $-49.48$
    & $0.55$ & $1.17$ & $0.51$ \\
    
    $\pi/4$  
    & $90.00$ & $39.58$  & $-12.61$ & $45.00$ & $-57.61$ & $45.00$  
    & $90.00$ & $112.50$ & $84.38$  & $-45.00$& $-84.38$ & $45.00$ 
    & $0.49$ & $1.05$ & $0.69$ \\
    
    $\pi/8$  
    & $90.00$ & $-158.36$& $27.91$  & $45.00$ & $62.09$  & $-45.00$ 
    & $90.00$ & $-78.75$ & $87.19$  & $-45.00$& $-87.19$ & $45.00$ 
    & $0.00$ & $1.05$ & $0.90$ \\
    
    $0$      
    & $90.00$ & $180.00$ & $-22.50$ & $45.00$ & $-67.50$ & $-45.00$ 
    & $90.00$ & $90.00$  & $-90.00$ & $-45.00$& $-90.00$ & $45.00$ 
    & $0.00$ & $1.00$ & $1.00$ \\
    \bottomrule
  \end{tabular}
  }
  \caption{
    Optimal local general-axis measurement parameters $(\beta_i,\alpha_i)$ and
    implementing waveplate settings $(h_i, q_i)$ for varied graph weights
    $\phi_{12}$. Waveplate columns correspond to the QWP/HWP angles for the $(+)$
    or $(-)$ projector of each qubit. Also shown are the resulting minimum
    estimator variance $(\Delta \theta)^2$, expectation values $\expval*{\hat{A}}$,
    and derivatives $\abs*{\partial_{\theta}\expval*{\hat{A}}}$. Angles are
    reported in degrees.
  }
  \label{tab:general_axis_params}
\end{table}

\begin{figure}[h!]
    \centering
      \includegraphics[width=\linewidth]{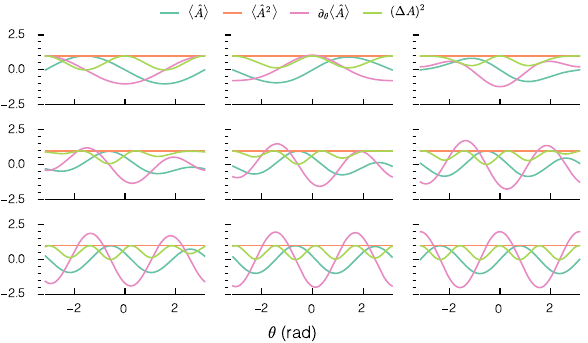}
      \caption{
	  Top left to bottom right: The expectation value of
	  $\expval*{\hat{A}}$ (blue line), its square $\expval*{\hat{A}^2}$
	  (orange line), derivative $\partial_{\theta}\expval*{\hat{A}}$ (green
	  line) and variance $(\Delta A)^2$ (red line) as a function of the
	  encoded phase $\theta \in [0, 2\pi)$ for graph weight $\phi_{12} \in
	  \{\pi, \tfrac{7\pi}{8}, \tfrac{3\pi}{4}, \tfrac{5\pi}{8},
	      \tfrac{\pi}{2}, \tfrac{3\pi}{8}, \tfrac{\pi}{4}, \tfrac{\pi}{8},
	  0\}$. Here, $\hat{A}=\hat{A}_{1}\otimes\hat{A}_2$ is the optimal
	  general-axis operator at the phase encoding $\theta = 0$. These
	  operators can be found in Tab.~\ref{tab:general_axis_params}.
      }
      \label{fig:general_axis_behaviour}
\end{figure}

\newpage
\noindent
Similar to the local Pauli observables, near the operating point $\theta=0$,
the derivative and variance of the local general-axis operators $\hat{A}$ are
robust to changes in $\theta$. Fig.~\ref{fig:general_axis_behaviour} shows the
expectation value of $\expval*{\hat{A}}$, its square $\expval*{\hat{A}^2}$,
derivative $\partial_{\theta}\expval*{\hat{A}}$ and variance $(\Delta A)^2$ of
the optimal local general-axis operators as a function of the encoded phase
$\theta \in [0, 2\pi)$ for different graph weights.

\end{document}